%% file: papercor.tex
\documentclass[a4paper, 12pt]{article}
\usepackage{epsfig}

\newcommand{\be}{\begin{equation}}
\newcommand{\ee}{\end{equation}}
\newcommand{\ba}{\begin{eqnarray}}
\newcommand{\ea}{\end{eqnarray}}
\newcommand{\bi}{\begin{itemize}}
\newcommand{\ei}{\end{itemize}}
\newcommand{\tr}{{\rm Tr\,}}
\newcommand{\re}{\mathop{\rm Re}}

\newcommand{\nn}{\nonumber \\}
\newcommand{\<}{\langle} %{\left\langle}
\renewcommand{\>}{\rangle}  %{\right\rangle}
\newcommand{\eq}{Eq.~}
\newcommand{\eqs}{Eqs.~}
\newcommand{\fig}{Fig.~}
\newcommand{\tab}{Tab.~}
\newcommand{\la}{\label}

\newcommand{\n}{$N$~}

\newcommand{\diag}{\mathop{\rm diag}}
\begin{document}
\begin{titlepage}
\begin{flushright}
%hep-lat/0509018\\
DESY 05-154
\end{flushright}
\begin{centering}
\vfill

{\Large{\bf Hot QCD, $k$-strings and the adjoint monopole gas model}}
\vspace{1.5cm}
  
\centerline{{\large Chris~P.~Korthals Altes}}
\centerline{\sl Centre Physique Th\'eorique au CNRS}
\centerline{\sl Case 907, Campus de Luminy, F13288, Marseille, France}

\vspace{0.8cm}

{\large Harvey~B.~Meyer}\footnote{harvey.meyer@desy.de}
\centerline{\sl DESY}
\centerline{\sl Platanenallee 6}
\centerline{\sl D-15738 Zeuthen}

\vspace*{2.0cm}

\end{centering}

\begin{center}
{\bf Abstract}
\end{center}
\noindent When the magnetic sector of hot QCD, 3D SU($N$) Yang-Mills theory,
is described as a dilute gas of non-Abelian monopoles 
in the adjoint representation of the magnetic group, 
Wilson loops of ${\cal N}$-ality $k$ are known to obey a periodic $k(N-k)$ law. 
Lattice simulations have confirmed this prediction to a few percent
for $N=4$ and 6. We describe in detail how the magnetic flux 
of the monopoles produces different area laws for spatial Wilson $k$-loops.
A simple physical argument is presented, why the predicted and observed 
Casimir scaling is allowed in the large-$N$ limit 
by usual power-counting arguments. 
The same scaling is also known to hold in two-loop perturbation theory
for the spatial 't Hooft loop, which measures the electric flux.
We then present new lattice data for 3D $N=8$ $k$-strings as long as 3`fm' 
that provide further confirmation.
Finally we suggest new tests in theories with spontaneous breaking
and in $SO(4n+2)$ gauge groups.

\noindent 
\vfill
%\noindent
%PACS numbers: 
%12.38.Gc, %        Lattice QCD calculations
%\\
%Keywords:
%lattice simulations

\vspace*{1cm}
\vfill
\end{titlepage}

\setcounter{footnote}{0}

%%%%%%%%%%%%%%%
\section{Introduction}
%%%%%%%%%%%%%%%
% high T
% power of 1/N correction?
% Casimir scaling. Diffusion on group manifold. Natural in models:
% dimensional reduction.
% magnetic monopole gas -> area law for temporal Wilson loop
% electric gluon gas    -> area law for spatial  Wilson loop at high T
% numerically, 3d is faster
%%%%%%%%%%%%%%%%%%%%%%%%%%%
The title of this paper may sound to most practitioners of lattice gauge theory
and hot QCD of a somewhat esoteric nature. And on the other hand 
aficionados of the beauty of non-Abelian monopoles~\cite{god,bais,bais98} 
\cite{abouel,olive80} \cite{brandt,konishi}
may reflect on the title as being heretic, since non-Abelian monopoles 
have so far withstood the traditional approach that has been implemented
successfully for 't Hooft-Polyakov monopoles: as yet, nobody has 
come up with a viable construction of a classical solution, 
that is then quantized by semi-classical methods.
Only recently a construction of non-Abelian fluxes 
in a low energy field theory version has been accomplished~\cite{hanany}. 
These are models that relegate the intricacies of non-Abelian monopoles 
to their high energy sector, and manage to construct explicit 
non-Abelian \emph{fluxes} in the low energy sector.

Somewhat analogously, we will forget about the intricate nature of individual non-Abelian 
monopoles 
and assume that a \emph{gas} of such objects has relatively straightforward properties. 
That will allow us to compute and interpret in a simple-minded way the average behaviour
of magnetic flux loops, that is, spatial Wilson loops~\cite{giovanna01}. 
Such spatial loops have been measured in lattice simulations
by Teper's group~\cite{lucini,lucini04} in a wide temperature range, thus showing
that the predictions of the model can be tested from first principles.

At temperatures well above the critical $T_c$, 
the temporal extent of the system becomes negligible, 
and we are left with a three-dimensional system. 
This implies that the tension of the spatial loop at such very high $T$ 
also bears the interpretation of a three-dimensional string tension, 
due to a chromo-electric flux tube. In other words, 
%provided the non-Abelian monopoles survive the dimensional reduction, 
our model is also 
indirectly a model for confinement in 2+1 dimensional gauge theories.

Generally speaking, in non-Abelian SU($N$) gauge theories 
in three and four dimensions, the 
chromo-electric flux between two static colour sources 
arranges itself so as to produce a linearly rising potential.
This naturally suggests a flux-tube configuration and leads to the 
string picture of confinement. While in SU(3) there is only one
string tension, that of the string appearing between charges 
in the fundamental representation, in SU($N\geq 4$) there are $[N/2]$
independent stable `$k$-strings' which are protected from screening 
by the center-symmetry $Z(N)$.

The picture that we propose for the origin of the area laws of the spatial 
Wilson $k$-loops, and hence for 3d $k$-strings, is rooted (perhaps paradoxically)
in high temperature 3+1 dimensional QCD and involves a gas of screened
non-Abelian monopoles -- or rather ``magnetic quasi-particles''. 
We prefer the latter terminology, since it stresses that our monopoles need  
not be eigenstates of the Hamiltonian but are rather collective modes of the plasma.
The objects that we shall describe in the 3d gauge theory are the dimensionally reduced 
versions of these modes, much in the same way as Polyakov's 
`pseudoparticles'~\cite{polyakov} in the 3d Georgi-Glashow model
are the descendants
of the t'Hooft-Polyakov monopoles~\cite{th74} living in the 4d Georgi-Glashow model.
The non-Abelian Stokes theorem~\cite{diakonov} establishes a connection between
spatial Wilson loops and the magnetic flux in the plasma; which in our model 
is induced by the magnetic quasi-particles. That is, schematically, 
how we are able to make predictions for 3d $k$-string tensions.

Of course, $k$-strings are also interesting in their own right.
Since they are perfectly stable, their tension ratios can be used 
to discriminate unambiguously between models of confinement.  
In what follows, without giving a comprehensive view of the latter,
we put our model in perspective with respect to a broader class of such models.

Our adjoint monopole gas model~\cite{giovanna01,altes04} 
is related to the dual-superconductor picture of confinement~\cite{tH76}.  
The latter would naturally predict the presence
of monopoles in the plasma, as manifestations of the condensate at low $T$.   
It is a natural generalization of the seminal idea of 't Hooft~\cite{thooft80}, 
that Abelian monopoles Bose-condense in the ground state, and are transient 
states in that they won't show up in the spectrum of the Hamiltonian. In the 
hot deconfined phase they should populate the ground state, just like gluons.
To explain the $k$-loop tensions in the hot phase is however non-trivial
because the number of different species of Abelian monopoles is too small 
($N-1$ for SU($N$)).

There is the elegant caloron solution to the equations of motion~\cite{vanbaal}. 
It is a periodic instanton with a Higgs-like background furnished by the non-trivial
value of the Polyakov loop. This gives rise to $N$ monopoles (a fundamental multiplet) 
inside the caloron. Could these be related to the quasi-particles that we are invoking?
It may be~\cite{vanbaal} that at high enough temperatures the
monopoles inside an individual caloron start to ``deconfine'' and are 
able to move freely from one to another caloron, much in the same vein that gluons
can freely move from one glueball to another at high $T$. However free monopoles in 
the fundamental multiplet can \emph{not} explain the observed Casimir 
scaling~\cite{altes04}. Nevertheless, as explained in the next section, 
even at asymptotic temperatures we are actually facing 
\emph{strong} coupling when we try to explain the spatial Wilson loop behaviour. 
It could well be that this strong coupling favours binding into adjoint monopoles
(while binding into singlets is statistically disfavoured at large $N$). 
Non-Abelian monopoles in the \emph{adjoint} representation furnish
precisely the correct number of species to explain the observed
Casimir scaling, as shown in earlier
work~\cite{giovanna01} and in section \ref{sec:effect} below.

The ratios of $k$-string tensions are also tests for formulations 
of SU($N$) gauge theories derived from fundamental string theory.
Examples of the latter are the MQCD framework~\cite{kleb} and the
AdS/CFT calculations in Ref.~\cite{herzog} for $D=3$. The latter give
Casimir scaling for $N$ large and $k$ of order $N$; 
our model predicts Casimir scaling for any value of $N$.

The MQCD framework gives a $\sin(k\pi/N)$ law for the k-tension, 
implying in particular that the tension ratios $\sigma_k/\sigma_1$ 
have $1/N^2$ corrections. 
An elegant paper by Gliozzi~\cite{gliozzi} provides a simple geometric 
interpretation for the sine law. He shows that in the cold phase the sine law 
is the borderline between formation of $Z(N)$ symmetric static baryons 
(no $k\ge 2$ flux tubes involved) and formation of static baryons with
$k=2$ or higher flux tubes (it is assumed that arbitrary short
flux tubes have the same tension as long flux tubes).

Casimir scaling and the sine law both predict that 
$\sigma_k/\sigma_1\to k$ at large $N$, fixed $k$; in other words,
a $k$-string is a collection of $k$ non-interacting fundamental 
strings in the planar limit $N\to\infty$.
Casimir scaling however attributes a binding energy to these 
$k$ strings of order $1/N$, while if the sine law is correct,
this energy is only ${\cal O}(1/N^2)$.
Recently there has been a discussion~\cite{armoni} on the conflict 
of $1/N$ corrections with standard $1/N$ power-counting rules, 
based on the assumption that \emph{all} 
representations with a given ${\cal N}$-ality $k$ produce the same tension. 
We point out in section \ref{sec:lattice} that this analysis neglects 
mixing effects between reducible representations which are of order $1/N$
and which lower the energy of the lightest string by an amount of that order. 
Earlier work on strong coupling expansions \cite{washington}
corroborates our general argument. More recently, analytic calculations 
of the tension for 't Hooft loops~\cite{thooftloop,giovanna01} have been
shown to lead to the same $k(N-k)$ scaling law.

As already mentioned, lattice calculations have been carried 
out~\cite{lucini,debbio,lucini04} in three and four dimensional SU($N$)
gauge theory to determine the ratios of the $k$-string tensions to the 
fundamental string tension. Here we study the $k$-strings in 3d
SU($N$) gauge theories, presenting new data for their tension ratios 
obtained for the gauge group SU(8) and combining the new information 
with previously obtained SU(4) and SU(6) data~\cite{lucini}.  
The numerical advantage of searching for the effect of monopoles 
on Wilson loops at high $T$ is that the 
relevant simulations are three-dimensional;
needless to say, to obtain the same accuracy, 
the amount of computational effort is considerably lower
for the 3d simulations employing the reduced action.

By the same token, given such an accuracy for the 3d lattice data, 
it is useful to know to what accuracy in the coupling $g(T)$ 
the dimensionally reduced actions reproduce the full 4d QCD result. 
For the case of three colours one knows~\cite{yorkmikko} 
that the 3d results for the string tension reproduce 
the 4d lattice data up to $1.1 T_c$ through the running 
of $g(T)$ up to and including two loops.

The lay-out of the paper is as follows. We start with  section \ref{sec:hotqcd} 
on how the problem of the residual strong interactions in hot QCD is attacked 
quantitatively -- by dimensional reduction. In section \ref{sec:namon} we review
briefly non-Abelian monopoles. We then derive in section \ref{sec:stokes}
a Stokes type formula for the spatial Wilson loops that permits us to
quantify the effect of the putative non-Abelian monopoles in section
\ref{sec:effect}. Then follows section \ref{sec:string_high_rep} on
strings in higher representations where our arguments on the $1/N$
counting are exposed, and the lattice calculation is
presented in section \ref{sec:lattice}. Finally we compile and discuss 
the lattice data accumulated so far (section \ref{sec:discuss}) 
and the paper ends with a general conclusion (section \ref{sec:conclu}).

%%%%%%%%%%%%%%%%%%%%%%%%%%%%%%%%%%%%%%%%%%%%%%%%%%%%%%%%%%%%%%%%%%%%%%%%%%%%%%%%%%%%%%
\section{High temperature QCD}\label{sec:hotqcd}
%%%%%%%%%%%%%%%%%%%%%%%%%%%%%%%%%%%%%%%%%%%%%%%%%%%%%%%%%%%%%%%%%%%%%%%%%%%%%%%%%%%%%%
This section is meant to introduce the reader into the essentialia of hot
QCD, and to motivate the model.

At temperatures well above $T_c$ asymptotic freedom drives the running 
coupling $g(T)$ down to zero.  
On the other hand the average density of gluons is the 
Bose-Einstein density $n_{BE}(p/T)$ ($p=|\vec p|$ is the momentum of a gluon). 
Because of this density the coupling 
in the plasma has to describe stimulated emission and equals
\be
g^2_{st}=g^2n_{BE}(p/T)=g^2{1\over{\exp{p/T}-1}}. 
\ee

This leads to a picture of a gluon plasma,
where one has to distinguish three scales:
\begin{itemize}
\item{hard gluons with momentum $p$ of order T, interacting weakly, 
$g^2_{st}={\cal O}(g^2)$.}
\item{soft gluons with momentum $p$ of order $gT$, still interacting weakly
 $g^2_{st}={\cal O}(g)$.}
\item{ultra-soft gluons with momentum $p={\cal O}(g^2T)$, interacting strongly,
 $g^2_{st}={\cal O}(1)$.}
\end{itemize}

Thus, in spite of asymptotic freedom, there is a strongly interacting sector left.
Strongly interacting because the large population of ultra-soft energy levels
pushes the coupling up~\cite{linde}. Thus, at these length scales, semi-classical
methods are unlikely to apply, as we argued in the introduction.

The hard gluons are familiar from the Stefan-Boltzmann form for the pressure.
The  hard gluons cause Debye screening $m_D\sim gT$ of the force between electric
test charges. All this has been known for long from electrodynamic plasmas.

A new feature becomes apparent for the non-Abelian plasma at scales $g^2T$. 
It is the screening of the magnetic force ($m_M\sim g^2T$) between two static 
magnetic test charges. In electrodynamic plasmas no static magnetic screening exists.
Magnetic screening not only occurs at arbitrary high temperatures, 
it {\it persists} at arbitrary low temperatures, where the electric screening 
has disappeared and has turned into electric confinement.  
It is a hallmark of the non-Abelian system, and  hints at a magnetic activity 
for {\it  all} temperatures~\cite{manuel}.

One purpose of this paper is to understand and test a specific model~\cite{giovanna01}
for the strongly interacting sector. We will state its assumptions
at the end of this section.

\subsection{Dimensional reduction at high $T$}\label{subsec:dimred}

In this section we give a fast review of how one computes equilibrium properties
of the plasma in a systematic way.  
The problem of strong coupling at large distances is dealt with through a
sequence of effective actions~\cite{pis}. It is the last and strongly
interacting effective (``magnetic'') action that our  monopole  model approximates.

By integrating out the hard modes in the QCD action one produces an effective 3d 
action called $S_{EQCD}$. If one accepts to have an accuracy of ${\cal O}(g(T)^4)$
this electrostatic action is the superrenormalizable action in terms of
the static potentials.
The form of our effective action $S_{EQCD}$ is dictated by all symmetries, 
global and local, of the original QCD action, which are respected 
by the integration process. That implies all the symmetries we knew already,
  except that the electric term in the static action will 
have no $\partial_0\vec A$ term. So $A_0$ appears as an adjoint Higgs 
term in our 3D gauge theory. The electrostatic QCD action density reads:
 \begin{eqnarray}
{\cal{L}}_{E} & = & \tr\{(\vec D(A)A_0)^2\} + m_E^2\tr\{A_0^2\} + 
                \lambda_E(\tr\{A_0^2\})^2+ {} \nonumber\\
              & + &  \bar\lambda_E \left[ (\tr\{A_0\})^4-{1\over 2}(\tr\{A_0^2\})^2\right] +
               {1\over 2}\tr\{ F_{ij}^2\} + \delta{\cal{L}}_E.
\label{eq:estat} 
\end{eqnarray} 
Because of R- conjugation invariance ($A_{0}\rightarrow -A_{0}$) 
the electrostatic action must be even in $A_0$.  For SU(2) and SU(3)
the second quartic term is identically zero.

The parameters in this 3d action are  the coupling $g_E$, the electric mass
$m_E$ and the 4-point couplings $\lambda$.  All of them are expanded 
in powers of the QCD running coupling $g^2(T)$, and all of them are now known to 
${\cal O}(g^4)$~\cite{huang,kaj1997}. The electric mass coincides with 
the Debye screening mass $m_D^2={g^2N\over 3}T^2$ to one loop order. 
The 4-point couplings start with the fourth power of $g(T)$.
It is customary~\cite{kaj98} to express all these parameters in terms of the 
dimensionful scale $g_E$, $x=\lambda_E/g^2_E$, similarly for $\bar x$, and finally
$y=m_E^2/g_E^4$. For large $T$ the $x$ variable becomes equal to $g^2$, the $xy$ 
variable approaches a constant.

In the limit where the electric mass $m_E\sim gT$ becomes very  large compared to the 
coupling $g^2_E=g^2T$ one can integrate out this mass scale and obtain magnetic QCD:
\begin{equation}
{\cal{L}}_M={1\over 2}\tr\{F_{ij}^2\} + \delta{\cal{L}}_M.
\label{eq:mstat} 
\end{equation}  
The coupling parameter in this Lagrangian is called $g_M$ and it can be expressed 
in terms of the electric coupling $g_E$~\cite{giovanna03}: 
\begin{equation}
g^2_M=g^2_E\left[1-{g_E^2\over {16\pi m_E}}-{17\over{512}}
\left({g_E^2\over{\pi m_E}}\right)^2\right].
\label{eq:magnel}
\end{equation}

One should realize that in the pure $N=3$ Yang-mills theory there is only 
one parameter: the  coupling $g(T/\Lambda_T)$. This means there is a relation 
between $x$ and $y$, where the physics of the plasma is:
\begin{equation}
xy|_{4D}={3\over{8\pi^2}}\left[ 1+{3\over 2}x+{\cal O}(x^2)\right] ~~\mbox{for}~~ N=3.
\label{eq:nisthree}
\end{equation}

This action serves to compute the leading contribution to magnetic quantities
like the spatial Wilson loop $\sigma$, 
or the magnetic screening $m_M$ at very high $T$. For dimensional reasons both are
proportional to $g_M^2$.  The corrections are very small, 
and this seems to be a general feature of this type of corrections~\cite{giovanna03}. 
On the other hand the corrections due to hard modes in  two loop approximation 
are appreciable~\cite{yorkmikko}. 
It turns out that one can extrapolate the 3d result for the Wilson loop to
about 1.1$T_c$ just by using this  2-loop running of the coupling $g_E$,
as a very good approximation to the 4d results. 
Quite likely the same is true for the magnetic screening length $l_M$ or magnetic
screening mass $m_M=l_M^{-1}$, which is defined from the correlation of a 
heavy monopole pair: as for the spatial tension, its dominant contribution comes
from the 3d magnetic sector. 
%It can be shown~\cite{owealtesdefor} for \emph{all} temperatures to equal  
%the $0^{++}$ mass gap in a 3d Hamiltonian with one periodic  direction of length $1/T$. 
%Thus we know its value at T=0 and $T=\infty$ from lattice 
%data~\cite{lucini04,teper98}. Reliable data are still lacking in between, 
%but should be available soon.
%
%The dimensionless ratio $\delta=\sigma/m_M^2$ must then be 
%a constant, independent of $T$ in the plasma phase, which is determined by the 3d ratio. 
%This ratio is small for any number of colours, 
%on the order of 5\% at $T=\infty$ and 8 to 9\% for T=0. 
%This ratio will play an important role in the sequel, as it 
%is a measure for the diluteness of the monopoles~\cite{altes03}.

Summing up: computing magnetic quantities at $T\gg T_c$, in 3d
magnetostatic QCD, is sufficient to know them all over 
the deconfined phase by simply using the two loop running of the coupling. This means that  some salient features of our model for the magnetic sector (see next subsection) are valid for all of the deconfined phase.

\subsection{Magnetic quasi-particle model for the magnetic sector}\label{subsec:model}

The magnetic sector is governed by 3d Yang-Mills theory. 
For the physics over distances larger than the magnetic screening length 
we make three Ansaetze:
\begin{enumerate}
\item{The interaction for the magnetic gluons is so strong that they bind in lumps. }
\item{The lumps are dilute.} 
\item{The lumps are non-Abelian monopoles.} 
\end{enumerate}
Their size is on the order of the magnetic screening mass $m_M={\cal O}(g^2T)$. 
And so is their inter-particle distance, or their density $n_M$. 
The ratio of the two corresponding volumes is the diluteness 
\be
\delta=n_M/m_M^3.
\label{dilute}
\ee
As the coupling drops out in this ratio  there is no {\it parametric}
reason that the diluteness is small. 

From these Ansaetze follows from simple dilute gas arguments 
(repeated in section  \ref{sec:effect}) that the tension 
$\sigma$ of the spatial Wilson loop equals: 
\be
\sigma\sim {1\over{m_M}}n_M.
\label{tensiondilute}
\ee
There is a group factor in front (which will be discussed 
in section~\ref{subsec:othermulti}), 
and the the corrections are in powers-not necessarily integer- of the diluteness.  

So the diluteness is known, once the tension and the magnetic screening mass are
known from lattice measurements. Its smallness is a dynamical effect giving 
a value of about $0.05$ (with a correction of ${\cal O}(1/N^2)$~\cite{teper98})
in the hot phase for SU($N$) groups: it is given by  the string tension 
in the 3D gauge theory in units of the lightest glueball mass.
Clearly it is gratifying to have an -- admittedly empirical -- 
justification for the diluteness being small.

It is instructive to compare our dilute gas of composite lumps with radius $l_M$ to 
the dilute gas one finds in the usual weak coupling plasmas. There the lumps are 
point-like particles, and the Debye screening length $l_D$ is {\it large} with respect
to the inter-particle distance, i.e. $l_D^3n \gg 1$, the  weak coupling plasma condition.
For hard gluons one has $n\sim T^3$ and $l_D^{-2}\sim g^2$ and the plasma condition is fulfilled.

We want to close this section with a brief comment.
It is tempting to go down from infinite temperature to finite temperature,
and consider the lumps as magnetic quasiparticles, or collective excitations of
the plasma. From our theoretical knowledge of the magnetic screening~\cite{owealtesdefor}
we know that at $T=0$ the screening mass equals the lowest glueball mass in
the 4D gauge theory; and the spatial Wilson tension equals the string tension at $T=0$. 
Lattice simulations~\cite{lucini} 
find the diluteness at zero temperature, as given by  the string tension  in units of the lightest
glueball mass in the 4D gauge theory,  is
{\it still} small, on the order of $0.09$ for $N$ large! Thus our dilute gas stays
dilute when lowering the temperature. At some temperature $T_q$ the Bose-Einstein statistics takes over  
(where the ratio of magnetic screening and de Broglie thermal wave length $T^{-1}$ become 
on the same order). And, in the spirit of dual superconductivity,  BE-condensation is then
marking the transition to the confined phase. 

%%%%%%%%%%%%%%%%%%%%%%%%%%%%%%%%%%%%%%%%%%%%%%%%%%%%%%%%%%%%%%%%%%%%%%%%%%%%%% 
\section{Non-Abelian monopoles}\label{sec:namon}
%%%%%%%%%%%%%%%%%%%%%%%%%%%%%%%%%%%%%%%%%%%%%%%%%%%%%%%%%%%%%%%%%%%%%%%%%%%%%% 

The magnetic sector of hot QCD has magnetic lumps through the strong ($g^2\sim 1$) binding of 
magnetic gluons. Very specifically we do not have a Higgs field at our disposal 
to define the U(1) field strengths.
The question is whether other than 't Hooft-Polyakov~\cite{th74} monopoles can be
formed under such circumstances. The answer is not known to date. But if they are there they must obey a Dirac condition.

In 1977, Englert, Windey, Goddard, Olive and Nuyts~\cite{god}  analysed precisely such hypothetical monopoles
in an unbroken gauge theory, and formulated the generalized Dirac condition. 
This condition is the following. Let B be a matrix in the SU($N$) Lie-algebra.
Let the colour magnetic field $\vec B$ be given far away from the monopole
by:
\be
\vec B=g\vec r{B\over{4\pi r^3}}.
\label{fieldstrength}
\ee
The Dirac condition then reads:
\be
\exp{igB}={\bf 1}.
\label{dirac}
\ee
This condition has to be fulfilled  for any matter field that couples to
the gauge field. Obviously we can take B to be diagonal.
Note also that for U(1) we get the expected result $gB=2\pi n$.

For any simple Lie group one has the orthogonal set of  diagonal generators 
$\vec H=(H_1,\ldots,H_r)$, with $r$ the dimension of the Cartan subalgebra.
The remaining orthogonal generators are $E_{\alpha}=E_{-\alpha}^{\dagger}$.
The roots $\vec\alpha=(\alpha_1,\ldots,\alpha_r)$ are given by: 
\ba
[\vec H, E_{\alpha}]    &=& \vec\alpha E_{\alpha}\nonumber \\
 \left[E_{\alpha},E_{\beta}\right]
&=& (\vec\alpha +\vec\beta)E_{\alpha+\beta} \quad \mbox{if $\vec\alpha+\vec\beta$ is a root}\nonumber \\
\left[E_{\alpha},E_{\beta}\right]  &=& 0 \qquad\qquad\qquad \mbox{otherwise}\nonumber \\
\left[E_{\alpha},E_{-\alpha}\right] &=& \vec\alpha\cdot\vec H.
\label{lieal}
\ea
These definitions imply a common normalization $\tr\{E_{\alpha}E_{-\alpha}\}=\tr\{H_i^2\}$. 
In physics we are used to have it equal to $1/2$. 
                                    
We define now the coroots $\vec{\hat{\alpha}}={\vec \alpha\over{{\vec\alpha}^2}}$.   
In terms of those the group admits a set of SU(2) subgroups (like the familiar I,
U and V spin in SU(3)) for any root $\alpha$, denoted by ${\rm SU}(2)_{\alpha}$. 
One gets them  by projecting $\vec H$ on the coroots and using the $E_{\alpha}$.
More precisely:
\ba
\hat E_{\pm\alpha}&=&{1\over{|{\vec\alpha}|}}E_{\pm\alpha}\nonumber\\
\hat H_{\alpha}&=&{\vec\alpha\over{{\vec\alpha}^2}}\cdot\vec H.
\label{su2}
\ea 
Crucial is now that the matrices  $\hat H_{\alpha}$, being homogeneous in the roots
and the $H_i$, are  {\it independent} of the normalization of  the matrices $H_i$.
Hence they  have eigenvalues, which are pure numbers. What are those?

The commutation relations that normalize $H_{\alpha}$  follow from \eq\ref{lieal},
with the result:
\be
[\hat E_{\alpha},\hat E_{-\alpha}]=\hat H_{\alpha}.
\label{su2sub}
\ee
The weights of an irreducible representation are given by the eigenvalues of the
diagonal operators. So if the carrier vectors $v_{(k)}$ diagonalize the representation
we can define the weights $\vec w_k$ by  :
\be
\vec H v_{(k)}=\vec w_k v_{(k)}.
\label{weights}
\ee
Now a theorem on Lie algebras~\cite{humphries} tells us that $2\hat H_{\alpha}$ has 
integer eigenvalues on any irreducible representation, and hence from \eq\ref{weights}:
\be
2{\vec\alpha\over{{\vec\alpha}^2}}\vec w_k ~~~\mbox{is integer.}
\label{irrep}
\ee

So on  any irreducible representation the eigenvalues of $H_{\alpha}$ are (half)-
integer. This fact tells us that the magnetic roots $\vec b$ defined by
\be
B={4\pi\over g}\vec b\cdot\vec H
\label{mroots}
\ee
must lie on a lattice generated by coroots $\vec b={\vec\alpha\over{{\vec\alpha}^2}}$ 
if the magnetic strength B obeys the Dirac condition, \eq\ref{dirac}. 
%This condition was found by Englert and Windey in Ref.~\cite{god}.
%
\begin{figure}[!htb]
\begin{center}
\makebox[7.0cm]{\psfig{figure=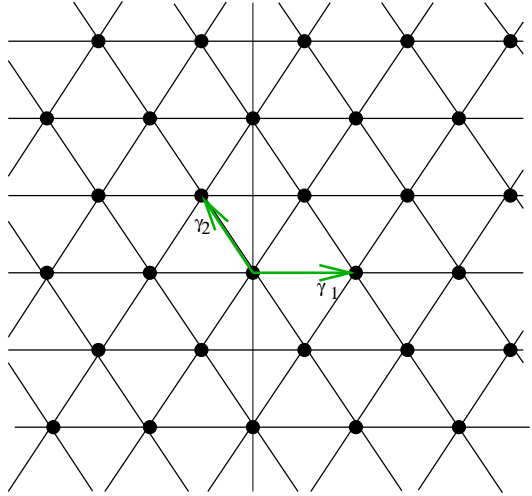,width=7.cm,angle=0}}
\caption[su3lattice0]{\footnotesize The lattice of
  allowed magnetic charges  spanned by the
  coroots for the gauge group SU(3).} 
\label{fig:su3lattice0}
\end{center}
\end{figure}
That is, every point $\vec b$ on the magnetic root lattice
is a linear combination with integer coefficients of
the coroots ${\vec\alpha\over{{\vec\alpha}^2}}$.

In \fig\ref{fig:su3lattice0} we show the coroot lattice for SU(3).
The two simple roots $\vec\gamma_{1,2}$ span the lattice. 
They define, with $\vec H$ in the fundamental representation, the matrices 
\ba
\vec\gamma_{1}\cdot\vec H\equiv H_{12}&={1\over 2}&\diag(1,-1,0)\\
\vec\gamma_{2}\cdot\vec H\equiv H_{23}&={1\over 2}&\diag(0,1,-1)
\ea
which indeed have half-integer eigenvalues and are the third components of I-spin
and U-spin respectively. Often it is more convenient to work 
with these matrices than with the root vectors themselves, as the former are truly simple.

For SU($N$), the simple roots are given by the generalization of I and U spin. 
The general representative is:
\be
H_{k,k+1}={1\over 2}\diag(0,0,...0,1,-1,0,...).
\label{simpleroot}
\ee
The first non-zero member is on the k-th diagonal entry, and $k$ ranges from 
1 to $N$, with:
\be
H_{N,N+1}\equiv H_{N,1}={1\over 2}\diag(-1,0,....,0,1).
\ee 
The sum of these matrices is zero, and usually the first $N-1$ are taken as 
simple roots.  It is then clear that we can rephrase the Dirac condition
as:
\be
B={4\pi\over g}\big(n_{12}H_{12}+....+n_{N-1,N}H_{N,N-1}\big),
\ee
where the $n$'s are 
integers\footnote{This notation is adapted to our notation for the Wilson loop in later sections.}.

This classifies the possible monopoles for all simple  classical Lie-algebras,
as hypothesized in the second paper of reference~\cite{god}.

For the group SU(2) the consequences of the Dirac condition and this
hypothesis are simple. We have a doublet with, in units of $4\pi/g$, $I_3=\pm 1/2$.
Then an iso-triplet with in the same units $I_3=\pm 1,0$. For a spin J 
(half)-integer multiplet we have the same. Our matrix $gB/4\pi$ with the 
spin 1/2 multiplet of magnetic charges gives only on integer spin electric
charge multiplets an integer. So the magnetic group of SO(3) is SU(2). 
On the other hand the iso-triplet of magnetic charges is compatible with any 
charge multiplet, half integer or integer, and so the magnetic group of SU(2) is SO(3). 

More generally, for the gauge group  SU($N$) all possible monopoles
are multiplets of a magnetic group ${\rm SU}(N)/Z(N)$. The opposite is also true:
the gauge group ${\rm SU}(N)/Z(N)$ admits monopoles in multiplets of the magnetic 
group SU($N$).

In \fig\ref{fig:su3lattice0} the lattice is shown for the gauge group SU(3);
for the gauge group ${\rm SU}(3)/Z(3)$ the lattice of monopoles 
will include the additional sublattice generated by the triplet representation. 
This additional sublattice
is obtained in a natural way by introducing the 2 hypercharges $Y_k$, k=1,2.  
Of course they are not uniquely defined.  They generate through
exponentiation $\exp(i2\pi Y_k)$ the center-group elements
of Z(3). We may for instance choose a set which is at a minimal distance 
(defined as the trace of the square of the matrix) of the center of the Cartan algebra:
\ba
Y_1&=&{1\over 3}\diag(2,-1,-1)\nn
Y_2&=&{1\over 3}\diag(1,1,-2).
\ea

In terms of the simple root matrices one finds:
\ba
Y_1&=&{2\over 3}(2H_{12}+H_{23})\nn
Y_2&=&{2\over 3}(3H_{23}+2H_{12}+H_{31}).
\ea 

So following  in \fig\ref{fig:su3lattice0} the steps along the weight lattice to
arrive at $Y_1$ one  gets the highest weight of the triplet representation. 
Similarly  $Y_2$ is the highest weight of the anti-triplet representation. 
In Appendix B we formulate this relationship for general SU($N$) and the generators
of its center-group $Z(N)$. Not only are the $N-1$ $Y$ matrices an alternative basis 
for the Cartan algebra. More important for us, they are a measure for the strength
of the Wilson loops needed to observe the monopoles (see section \ref{sec:stokes}) .

For any general classical Lie group it is the ``dual'' group~\cite{god}
built from the dual Lie algebra~\cite{humphries} that gives the possible multiplets.
The precise dual {\it group} with the appropriate center-group follows from the 
same considerations as for the SU($N$) case: the larger the original, electric, 
gauge group, the more stringent the Dirac condition becomes and a smaller magnetic
group follows. 
In mathematical terms it is the connectivity of the group and the ensuing Z(N) factors.

In an earlier paper~\cite{giovanna01}   precisely these hypothetical monopoles
were identified  with our lumps in 3 dimensions.  
As we supposed the lumps -- now monopoles -- to be dilute we can compute 
their effect on Wilson loops. For SU($N$) groups 
the choice of adjoint representation  is unquestionably favoured numerically, 
as simulated by $k$-loops for $N=4,~6$ by Teper~\cite{{lucini}} and for $N=8$ in this paper.

A comment  on the nature of the magnetic group is in order. The
monopoles, as bound states of magnetic gluons, will transform inside a
multiplet under some perhaps very complicated function of the original
vector potentials. So the global magnetic  SU($N$) group
will not  coincide with the original global colour group. This ties in
 with a phenomenon discovered by the authors in
Ref.~\cite{abouel}: global colour is not defined on the
quantized version of the non-Abelian monopoles, due to the long
range
nature of the colour magnetic fields of the monopole~\footnote{In the work by Bais
and collaborators~\cite{bais98} an interesting interpretation of the
magnetic group is proposed but its discussion falls beyond the scope
of this paper.}.

%%%%%%%%%%%%%%%%%%%%%%%%%%%%%%%%%%%%%%%%%%%%%%%%%%%%%%%%%%%%%%%%%%%%%%%%%%%%%% 
\subsection{Monopoles as a dilute gas: the broken symmetry case in the
            Georgi-Glashow model}\label{sec:ggmodel}
%%%%%%%%%%%%%%%%%%%%%%%%%%%%%%%%%%%%%%%%%%%%%%%%%%%%%%%%%%%%%%%%%%%%%%%%%%%%%% 

At this stage it is useful to put our model into a well-known context, 
the Georgi-Glashow model, with gauge group SU(2) in $D=3$ with gauge coupling $g_3$. 
According to our hypothesis we have a dilute gas of  iso-triplet monopoles
which describes the behaviour of Wilson loops.
Their density is proportional to $g_3^6$, the only scale before breaking the symmetry.
And their screening mass is proportional to $g_3^2$. Adding  an adjoint Higgs scalar
with a ``heavy'' VEV $v$, i.e. $v\gg g_3$,  will give us the broken phase 
with heavy 't Hooft-Polyakov monopoles:
\be
SO(3)\rightarrow U(1).
\ee

 This model was   studied by semi-classical methods in a seminal paper by 
Polyakov~\cite{polyakov}, in the limit that $g_3/v$ is small. 
In that limit the diluteness of the monopoles is a fact.
The Wilson loop tension $\sigma$ is exponentially small, 
like the density of the monopoles and the screening mass. 
The exponent is on the order of $\exp{-\gamma v/g_3}$, $\gamma$ 
some numerical constant. The result for the string tension can be expressed in terms
of the density of monopoles $n_M$ and the magnetic screening mass $M$ by combining:
\ba
\sigma& =& \frac{ g_3^2 M }{ 2 \pi^2 } 
\left[1 -{\pi M\over{ 2 g_3^2}} + \dots \right],\quad \mbox{and}\\
 n_M &=&  \frac{ g_3^2M^2 }{ 32\pi^2} \left[ 1+{\cal O}\left( {M\over{g_3^2}}\right)\right] 
\label{weakplasma}
\ea
into
\be
\sigma = \frac{16n_M}{M}  \left[ 1 + {\cal O}  \left( {M\over{g_3^2}} \right) \right]
\label{weaktension}
\ee
The correction to the tension is due to the authors in Ref.~\cite{jaimungal}.

\eqs\ref{weakplasma} and \ref{weaktension} are  typical for weak coupling plasmas.
The dimensionless ratio of tension over screening is proportional to 
the number of monopoles inside a sphere of radius the screening length:
\be
{\sigma\over {M^2}}=16{n_M\over{M^3}}\left[ 1+{\cal O}\left( {M\over{g_3^2}}\right) \right].
\ee

From \eq\ref{weakplasma}, this number is seen to be large: 
\be
{\sigma\over {M^2}}=16{n_M\over{M^3}}={g_3^2\over{ (2 \pi^2)M}}\gg 1.
\ee

In our model for the strongly interacting 
symmetric phase (see \eq\ref{dilute} and below),
this very same ratio is small! Physically, what happens is that 
the strong coupling creates a bound state of the original semi-classical monopoles
within the magnetic screening radius. And indeed,  as stated before, Monte-Carlo 
simulations in the symmetric phase give for the ratio ${\sigma \over M^2}$ 
0.046(2)~\cite{teper98}.

\subsection{Broken symmetry: SU(3)/Z(3) and higher groups}

Our next example is the gauge group ${\rm SU}(3)/Z(3)$ broken
by the adjoint to $U(1)^2$ or to  ${\rm SU}(2)\times {\rm U}(1)/Z(2)$.

\begin{figure}[!htb]
\begin{center}
\makebox[6.7cm]{\psfig{figure=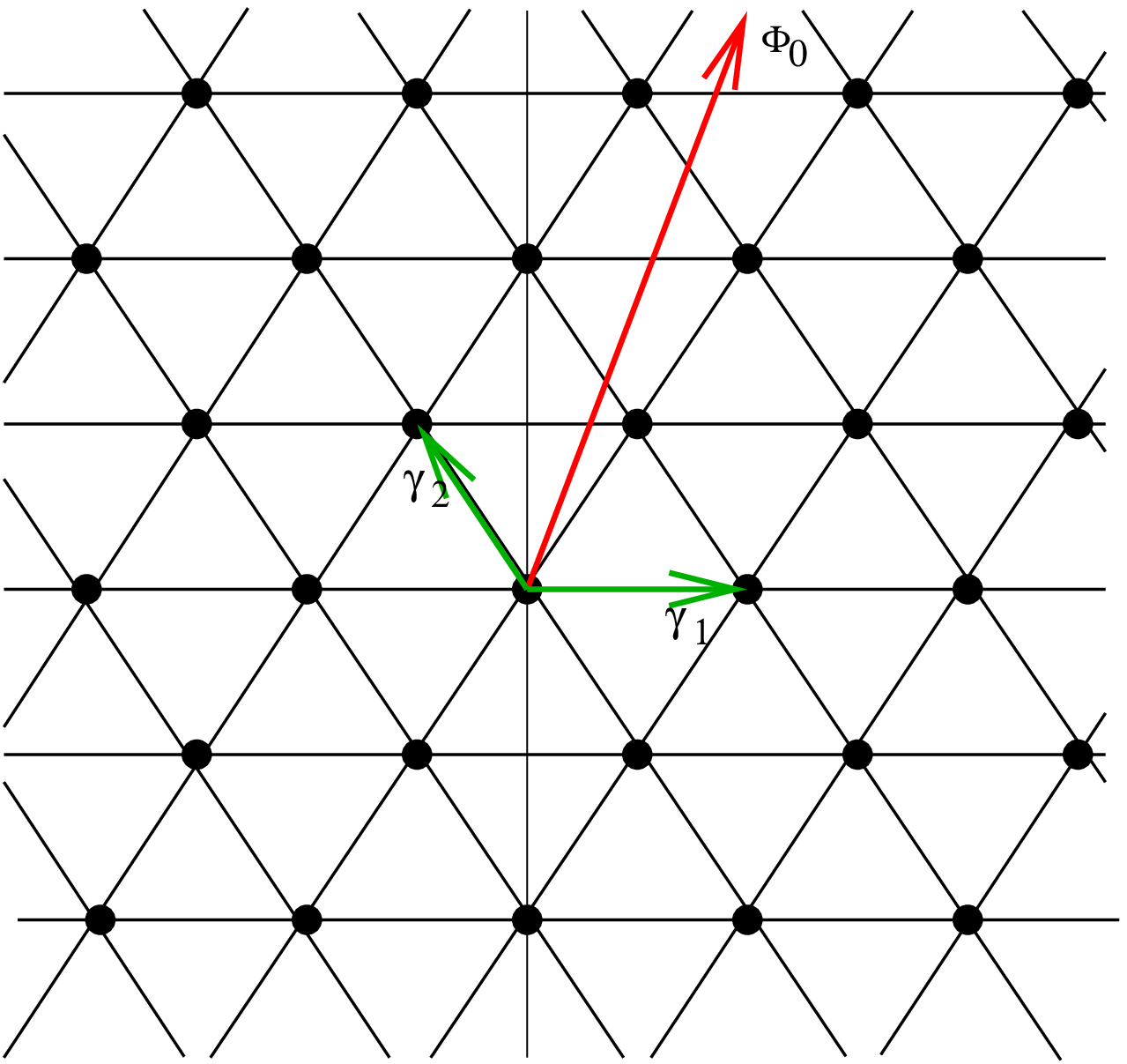,width=3.4cm,angle=0}}
\makebox[6.7cm]{\psfig{figure=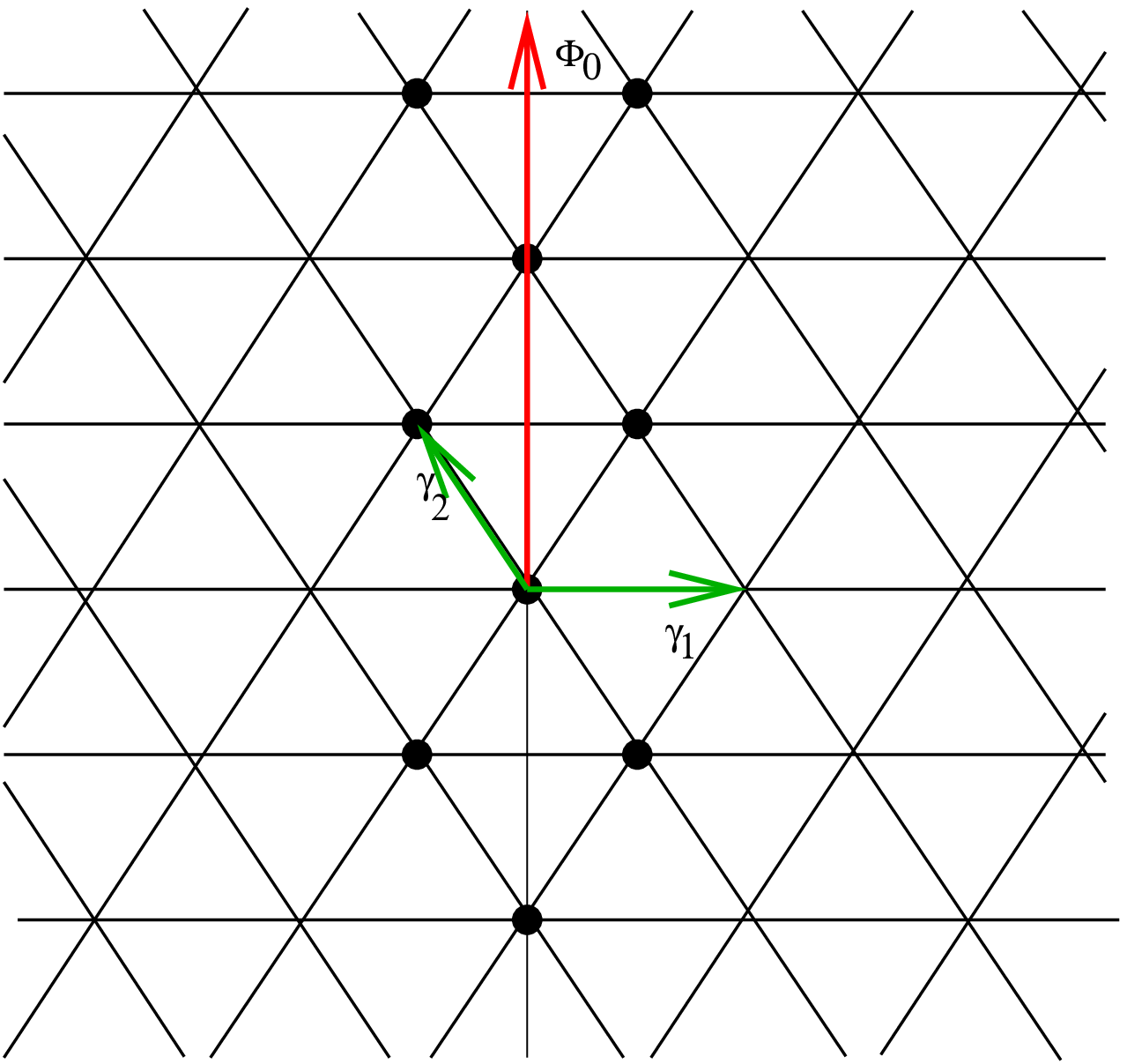,width=3.4cm,angle=0}}
%\makebox[3.7cm]{\psfig{figure=su3lattice3.eps,width=3.4cm,angle=0}}
 \makebox[6.7cm]{(a)} \makebox[6.7cm]{(b)}%\makebox[3.7cm]{(c)}
\caption[somethingelse]{\footnotesize
The lattice of charges
allowed by the quantisation condition, spanned by the
   simple coroots. In this figure also the direction of the Higgs field $\Phi_0$ in the
  Cartan subalgebra is indicated. In (a) the
  stable charges  for an arbitrary
  non-degenerate orientation of the Higgs field 
  are indicated by black dots. In that case the residual gauge
  group is $U(1)\times U(1)$ and all allowed charges
  correspond to a winding number.  In (b) the Higgs field is degenerate
  and leaves the non-Abelian group $U(2)$ unbroken. Now only one component of the
  magnetic charge is the winding number, and in each topological sector only
  the smallest total charge is conserved. The points symmetric with
  respect to the Higgs field are conjugate through the unbroken group.}
\label{fig:su3lattice}
\end{center}
\end{figure}

The first case is shown in \fig\ref{fig:su3lattice}.
Every point represents an 't Hooft-Polyakov monopole in the corresponding
SU(2) subgroup, as in \eq\ref{su2sub}. 
The Dirac condition carries integers which are the topological 
winding numbers of the Higgs field~\cite{bais98}.  
So this case does not go beyond what we already knew. 
As we wil see in the sequel this phase is not realized
in simulations, so we will not consider this phase anymore.

More interesting is the breaking pattern with unbroken group U(2).
This model is very often used \cite{konishi}  for 
investigations for non-Abelian monopoles. For momenta $p\gg v$ the broken phase 
is perturbative, for momenta much smaller than $g_3^2$ the coupling becomes strong. 
We expect screening at those distances, including screening of monopoles. 

If we try to construct an 't Hooft-Polyakov monopole in the  unbroken SU(2) 
group along $\vec\gamma_1$ in \fig\ref{fig:su3lattice}, we will fail because 
the VEV is lacking in that subgroup. Along the root $\vec\gamma_2$ the VEV is 
non-zero, so along that direction the integers still correspond to a winding number.
Similarly along the direction $\vec\gamma_1+\vec\gamma_2$, obtained by reflection 
of $\vec\gamma_2$ w.r.t. the direction of the Higgs breaking $\Phi_0$. Their long range 
magnetic fields are transforming into each other by the unbroken gauge transformations.
When trying to quantize these monopoles this property poses a problem~\cite{abouel}
of consistency, which is related to the fact that for the quantized solutions 
we expect screening. The long range field is unstable.

The mass of these objects is growing with the size of the VEV in the classical
approximation.  In what follows we will assume that this property survives 
the non-perturbative quantization. 
%%%%%%%%%%%%%%%%%%%%%%%%%%%%%%%%% FIGURE
\begin{figure}[tb]
\centerline{ %\hspace{-3.3mm}
\epsfxsize=9cm\epsfbox{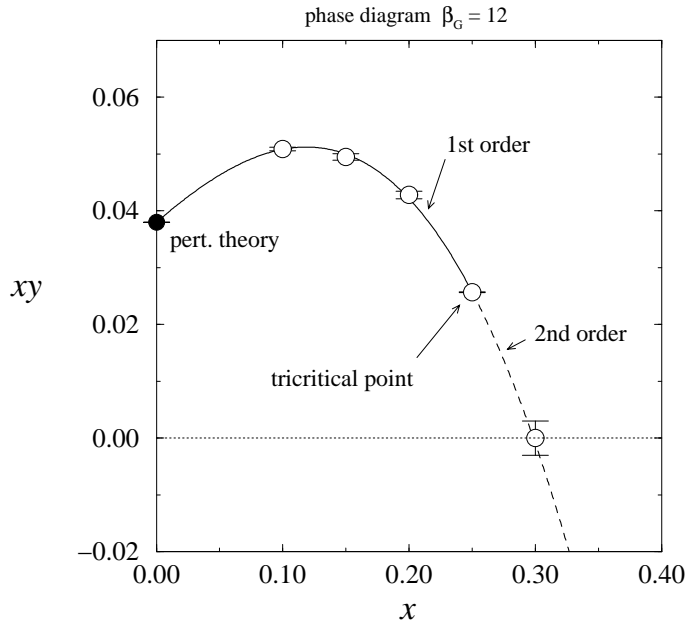}}
\caption[a]{The phase diagram of the 3d SU(3) + adjoint Higgs
theory~\cite{kaj98}.  The open symbols are results from the simulations, and
the filled circle is the perturbative result~\cite{bronoff,kaj98}.}
\la{fig:phasediag}
\end{figure}
%%%%%%%%%%%%%%%%%%%%%%%%%%%%%%%%%%%%
In terms of our model, the dilute gas of light octet monopoles in the symmetric phase will
after breaking leave the expected isotriplet in the unbroken SU(2), and two
heavy iso-doublets. The iso-triplet stays light after breaking, with a density $g_3^6$. 
The iso-doublets are the monopoles we described above and live at the lattice points 
$\vec\gamma_2$ and $\vec\gamma_1+\vec\gamma_2$ in \fig\ref{fig:su3lattice}. 
Due to their large mass they  have an exponentially small density like the 
't Hooft-Polyakov monopoles in the previous example.   
Note that in both examples the unbroken group defines a neutral singlet, 
that was present before breaking, 
but has disappeared in the transition between the two phases.

For illustration we show in \fig\ref{fig:phasediag} the phase diagram 
of the electrostatic theory given by the action in \eq\ref{eq:estat} 
for SU(3) by numerical simulation~\cite{kaj98}. The relevant variables 
are the dimensionless combinations $x$ and $xy$ discussed in section~\ref{subsec:dimred}.  
There is a first order  transition for small x, that is semi-classically 
calculable~\cite{bronoff,kaj98}. It marks the border of the  region 
where SU(3) symmetry is broken to ${\rm SU}(2)\times {\rm U}(1)$ and where the global R symmetry 
is spontaneously broken~\cite{bronoff}. For larger $x$ the transition becomes second order. 
Above the border there is the  unbroken phase. 
This unbroken phase can be smoothly accessed from the broken phase by the dotted, second order transition line. It means that 
putative monopoles in the unbroken phases are smooth deformations of the monopoles in the broken phase. 

Note the absence of a phase where $U(1)\times U(1)$ is unbroken but 
all other generators are broken~\cite{kaj98}. 
In general, for SU($N$), the adjoint Higgs does admit for 
breakings of the type ${\rm SU}(k)\times {\rm SU}(N-k)\times {\rm U}(1)$~\cite{rajantie}.

This phase diagram not only relates the putative monopoles in the unbroken 
phase to their more familiar analogues in the broken phase~\cite{rajantiealtes}.
In order to detect the monopoles one needs an operator that measures their flux.
This operator is the Stokes version of the spatial Wilson loop, and is intimately
related to a similar operator for the broken phase. This is the subject of the next section.   
 
%%%%%%%%%%%%%%%%%%%%%%%%%%%%%%%%%%%%%%%%%%%%%%%%%%%%%%%%%%%%%%%%%%%%%%%%%%%%%%%%%%%%%%%
\section{Flux representation for spatial 't Hooft and Wilson loops}\label{sec:flux}
%%%%%%%%%%%%%%%%%%%%%%%%%%%%%%%%%%%%%%%%%%%%%%%%%%%%%%%%%%%%%%%%%%%%%%%%%%%%%%%%%%%%%%%

The monopoles  have an effect
on spatial Wilson loops, because the loops record magnetic flux.
The traditional representation of the loop as a line integral is 
not appropriate to quantify the effect, and we have to find a flux
representation for the loop. For a loop in the U(1) case we have 
Stokes' theorem:
\be 
\exp{ig\oint_Ld\vec l\cdot\vec A}=\exp{ig\int d\vec S\cdot\vec B}\qquad\mbox{for U(1).}
\label{abelianstokes}
\ee

Now the non-Abelian case. For a certain class of irreducible representations 
 of SU($N$) one finds a  simple and useful generalization of the Abelian case.
It is   due to Diakonov and Petrov~\cite{diakonov}. 
If R is any one of the fully anti-symmetric  irreducible representations  
given by the one column Young tableau with $k$ entries it has highest weight $Y_k$ 
(see appendix A; recall the property $\exp{(i 2\pi Y_k)} = e^{i2\pi k/N}$).
Then  one finds for the Wilson loop $W_R(L)$:
\be
\tr {\cal P}\exp ig\oint_Ld\vec l\cdot\vec A_R =
\int D\Omega~\exp ig\int d\vec S \cdot \tr\left\{\Omega Y_k\Omega^{\dagger}\vec B\right\}.
\label{wilsonflux}
\ee
The integration is over {\it regular} gauge transforms $\Omega$.

In this section the physical ideas behind this Stokes law will be expounded. 
There are many  papers \cite{gubarev}  concerning its derivation, 
but we have not seen any exploring the significance of the class of 
gauge transformations involved, nor the special role played by the 
fully anti-symmetric irreducible representations. 
First we will make the Stokes theorem plausible by recalling some known 
features~\cite{kovneraltes} of colour electric analogue 
of the spatial Wilson loop: the spatial 't Hooft loop.

\subsection{Flux representation of the spatial 't Hooft loop}\label{sec:thflux}

Colour electric flux is confined inside glueballs. It is
only above the critical temperature that it becomes visible through the area law obeyed by the thermal average of the spatial 't Hooft loop.

The `t Hooft loop is defined as a loop of a Dirac vortex, 
with strength $z_k=e^{i2\pi k/N}$ in the center group. The vortex is 
created by a gauge transformation $\Omega_k$ with a discontinuity $z_k$, 
when circumnavigating the vortex. 
The locus of the discontinuity is a surface S spanned by the vortex L.

We take the simplest  gauge transformation that does have a discontinuity of this type.
Gauge transformations are generated by Gauss' operator $\vec D\cdot \vec E$.
If $\theta(S)$ makes a unit jump when going through the surface in the 
direction of the normal $\vec n$, then 
\be
V_k(L)=\exp i { 4\pi\over g} \int d\vec x~ \tr\left\{ Y_k(\vec E\cdot\vec D) \theta(S)\right\}
\ee
has the required discontinuity.

On physical states, only the gradient in the covariant derivative $\vec D=\vec \partial+ig[\vec A,$
counts~\cite{kovneraltes}. The gluon charge $gf_{abc}\vec A.\vec E$ is continuous through the surface. So the spatial 't Hooft loop becomes:
\be
V_k(L)=\exp i{4\pi\over g}\int d\vec S \cdot \tr \left\{Y_k\vec E\right\}
\qquad\mbox {(on physical states).}
\label{eflux}
\ee
This operator does not look gauge invariant, although on the physical subspace it is. In order to bring it in a manifestly
gauge-invariant form, we multiply it on the left with
a regular gauge transformation $\Omega$, and on the right with $\Omega^{\dagger}$;
a matrix element of $V_k$ between two physical states is not affected 
by this operation. After integration over all regular transformations,
\be
V_k(L)=\int D\Omega~\exp i {4\pi\over g} \int d\vec S\cdot \tr\left\{\Omega 
Y_k\Omega^{\dagger}\vec E\right\}
\qquad\mbox{(on physical states).}
\ee
is manifestly gauge-invariant.

It seems plausible to obtain its magnetic analogue by  replacing $\vec E$ by $\vec B$ and 
the coupling $\alpha \equiv g^2/4\pi$ by $\alpha^{-1}$.
That gives the formula for the Wilson loop, \eq\ref{wilsonflux}.  

\subsection{The flux representation for the Wilson loop}\label{sec:stokes}

The plausibility argument from the preceding section gives a formula
which is consistent with the expression given in Ref.~\cite{diakonov} for 
Wilson loop in any representation $R$, with highest weight $H_R$. 
Let $\Omega$ be any gauge transformation that is periodic on the loop. 
Then, with $\vec\nabla\Omega^{\dagger} =(\vec\partial-ig\vec A)\Omega^{\dagger}$ 
and $\vec\nabla\Omega=\vec\partial\Omega+ig\Omega \vec A$:
\begin{equation}
W_R(L)=\int D\Omega~\exp ig\int d\vec S
\cdot \tr\left\{[H_R\left(\Omega\vec B\Omega^{\dagger}-{1\over {ig}}
\vec\nabla\Omega\times  \vec\nabla\Omega^{\dagger}\right)\right\}.
\label{diakonov}
\end{equation}
This result, proved in Appendix B, 
differs from that of the plausibility argument through the presence 
of the second term. This term reduces in the SU(2) case to the familiar 't Hooft 
source term. If we  limit ourselves to
regular gauge transformations, this second term would drop out in the equations of motion.  

In the light of this we feel it is justified to 
make the following {\it assumption}: in our model with a dilute monopole gas 
the contribution of the second term is negligible.
A second simplification occurs when we are only interested in 
its  (thermal) expectation value.  
This is because in this case $W_R$ acts on the left and on the right only  on physical
states , so the effect of the regular gauge transforms $\Omega$ is undone.

There is a further comment related to this Stokes formula . 
It is derived under the assumption (see Appendix B) that it is regulated by the SU($N$) 
asymmetric top~\cite{diakonov}. The question is  whether the pure Yang-Mills theory average can be
provided with  such a regulator. For $N=2$ and in three dimensions the answer
is affirmative~\cite{kovneraltes} by adding an adjoint Higgs system and letting 
the VEV go to zero, followed by decoupling the Higgs in the infinite mass limit.
The VEV is the moment of inertia of the {\it symmetric } top. We can {\it not} 
accommodate the extra parameter of the asymmetric top, and this is the reason that 
the Stokes formula is then only valid for the fully antisymmetric irreducible representations with highest weight $H_R=Y_k$. 
The reason for this is that the second order Casimir operator takes its minimal 
value -- with  fixed {\cal N}-ality $k$ -- in the fully antisymmetric representation.

For general $N$ the answer is analogous, but Nature realizes only a limited set of
Higgs phases with only one adjoint Higgs field. They are limited to breakings of
the type where ${\rm SU}(k)\times {\rm SU}(N-k)\times {\rm U}(1)$ is still unbroken, 
$k\le [N/2]$~\cite{rajantie}.
That implies once more that the proof of the Stokes formula is only valid for 
those highest weights that have this symmetry, i.e. of the form $H_R=lY_k$, 
$l$ a positive integer. For $l=1$ this is the weight of the totally antisymmetric 
Young tableaux with $k$ boxes. Appendix B shows that $l>1$ is excluded.  
This ends our discussion of \eq\ref{wilsonflux}.

\subsection{Electric flux loop and its expectation value}

Let us return to the electric flux loop, \eq\ref{eflux}.
In this case the $\Omega$ integration drops out when one
acts with $V_k(L)$ on a physical state, because the only effect of
$V_k(L)$ is to multiply intersecting Wilson loops in the physical state
with a center-group factor (see \cite{kovneraltes} for more details):
\be 
\exp(i2\pi\Omega Y_k\Omega^{\dagger})=\Omega~\exp(i2\pi Y_k)~\Omega^{\dagger}=\exp(ik2\pi/N).
\ee

The thermal  expectation value of the 't Hooft loop has been calculated 
analytically at high temperature in powers of $g(T)$, including $g(T)^3$. 
This is possible because the effective potential is
in low orders built up by hard modes ({\cal O}(T)) and soft modes
({\cal O}(gT)). The ultra-soft magnetic modes come in at higher
orders. This potential has a $Z(N)$ symmetry, and the thermal
expectation value of the loop tension,
\be
\<V_k(L)\>= \exp\left( -\rho_k(T)A(L)\right),  \label{defetension}
\ee
is obtained from the  tunneling between two vacua, one corresponding
to $k=0$ and one corresponding to $k$. One finds then~\cite{thooftloop} that:
\be
\rho_k(T)=\rho_1(T){k(N-k)\over {(N-1)}},
\label{electrickscaling}
\ee
up and including two loop order. Concretely,
in one loop order the value of $\rho_1(T)$ is~\cite{bhatta}:
\be
\rho_1(T)={4\pi^2\over{3\sqrt{3g^2N}}}(N-1)T^2.
\label{electric1tension}
\ee
In three loop the above Casimir scaling is slightly invalidated, 
as it is found to be in lattice simulations~\cite{lucinidefor04,teper05}.

\subsection{Dilute gas approximation for both electric and magnetic flux loops}\label{sec:effect}
 
From the formulae in the preceding subsection one can easily find the behaviour 
of the tensions in terms of $k$, once one assumes a dilute gas of gluons for the
 electric loop and a dilute gas of monopoles for the magnetic loop at high temperature.
A dilute gas of gluons at high temperature will certainly disorder the electric loop. 
The reason is that the flux from one single species of gluons is going through 
the loop only when within screening distance $l_E$ from the loop:
\be
l_E \equiv \frac{1}{m_D} =  \sqrt{\frac{3}{g^2N}} ~\frac{1}{T}.
\label{debye}
\ee

Thus all the flux through the loop can only come from the gluon being in a slab 
of thickness $l_E$ and area $A(L)$ of the loop. Thus the flux is approximated
by a theta-function in the distance $d$ from the loop. 

The total flux from a charged gluon is $\pm 1$, as follows from the adjoint representation
 of $Y_k$. Thus the height of the theta-function is $\pm {1\over 2}$, because 
half of the flux is lost on the loop. Its effect on the loop is that it picks up  a factor
\be
V_k(L)=\exp \left[i2\pi \left(\pm {1\over 2}\right)\right] = -1.
\label{onegluon}
\ee
Note that not the value of the charge, but only the multiplicity of the charge
depends on $k$. This multiplicity is $k(N-k)$ for each value.

Now the distribution function of say $\ell$ gluons of a given charged species in such 
a slab is peaked around $\bar\ell$, the mean number of gluons in the box. Its width should, 
according to thermodynamics, be proportional to $\bar\ell$, like e.g. the Poisson distribution:
\be
P(\ell)={ e^{-\bar\ell}  \over{\ell!}}(\bar \ell)^\ell.
\ee
The average of the loop is therefore:
\be
\<V_k(L)\>|_{\rm one~species} = \sum_\ell P(\ell)(-1)^\ell=e^{-2\bar\ell}.
\label{averageonegluon}
\ee
Together with \eq\ref{defetension} this means that a single charged gluon species 
will determine the thermal average of the loop to be an area law:
\be
A(L)\rho_k|_{\rm one~species} = 2\bar\ell = 2A(L) l_E~ n(T).
\label{onespecies}
\ee
Note the absence of $k$ dependence in the outcome! What counts
is that the charge is non-zero, but its sign is 
irrelevant\footnote{The reader might be alarmed by our cavalier treatment 
of the screening of the flux. The flux $\Phi(d)$ that a gluon 
at distance $d$ shines through the loop
is exponential in $d$, not a theta-function! One can correct
for this by dividing the space above and below the loop in 
parallel slabs of infinitesimal thickness. This means the summand
in \eq\ref{averageonegluon} is replaced by an integral 
$\int d(d) \exp[1-\cos\{2\pi\Phi(d)\}]$. As a result  the factor
2 in \eq\ref{onespecies} increases by a factor 1.64282....}.

Thus the only way the $k$ dependence comes in is  when we take all the charged gluons
into account. This number, the multiplicity with respect to the charge $Y_k$,
is for the adjoint gluon multiplet equal to $2k(N-k)$. It is the number 
of non-zero entries in the diagonal adjoint representation of $Y_k$.  
We supposed the gluons to be independent; it follows that
\be
\rho_k|_{\rm all~species}=2\bar \ell \cdot 2 k(N-k)= 4 l_E~ n(T)~k(N-k).
\label{allspecies}  
\ee
Thus the $k$-loop is proportional to the multiplicity of charged gluons with respect 
to the charge $Y_k$. Note also that equation \ref{debye}, together
with the density of the gluons being $\sim T^3$, makes the 
outcome of the one species calculation  parametrically identical to the analytic
result in \eq\ref{electric1tension}.

The calculation of the magnetic loop is identical. The unit of magnetic charge 
is $4\pi/g$ instead of $g$,  but this is canceled by changing from electric to magnetic loop.
It is useful to realize that the surface integral $\int d\vec S\cdot \vec B$
for a single magnetic quasiparticle is given by ${1\over 2} B$, where $B$ is the 
magnetic charge matrix satisfying the Dirac condition (see section~\ref{sec:namon});
this condition thus directly leads to the phase $\pi$ necessary to disorder the Wilson loop,
for any member of the adjoint representation.
So the thermal expectation for the magnetic $k$-tension is, as for its electric counterpart:
\be
\sigma_k\sim l_M n_M(T) ~ k(N-k).
\label{magneticktension}
\ee
Though superficially very alike, there is an important difference between
the two tensions in units of the respective screening lengths.
The electric tension in those units becomes
\be
l_E^2\rho_k\sim l_E^3 n.
\ee
On the right hand side we have a large number, ${\cal O}(g^{-3}(T))$, for high T. This is
the plasma condition. It says that an electric screening volume contains a large
number of  almost free gluons. And corrections are in terms of inverse fractional
powers of this ratio, as discussed already in section \ref{sec:ggmodel}.
On the other hand the Wilson k-tension equals:
\be
l_M^2\sigma_k \sim l_M^3 n_M.
\ee
Both the magnetic screening and the magnetic density are ${\cal O}(g^2T)$.
So in the ratio the coupling drops out. Lattice data tell us the l.h.s. is small 
for all $N$.  The corrections are discussed in section \ref{sec:corrections}.

For large $N$ the dimensionless quantity $n_Ml_M^3$ is of order $1/N$. This is so because
the magnetic screening length $l_M$ can be shown to be given by the
$0^{++}$  mass of the Hamiltonian of 2+1 dimensional Yang-Mills theory, 
and therefore behaves parametrically like $1/g^2NT$. 
The density of a single monopole species $n_M$ 
should be ${1\over N}(g^2NT)^3$, in order to recover a tension of ${\cal O}(1)$.

\subsection{Monopole multiplets other than the adjoint}\label{subsec:othermulti}

Now  we use a general multiplet R carrying a unitary representation 
$D_R$of the magnetic group as the
magnetic quasi-particles in our model~\cite{altes04}. Its dimension is
$d_R$. The Lie- representative of the charge $Y_k$ is written as $(Y_k)_R$ 
and the corresponding group element as $D_R(Y_k)$.  Of course $D_R(Y_k)= \exp{i(Y_k)_R}$  

As in the previous subsection, the quasi-particle model produces for a given member 
$r$  ($r=1,2,\dots,d_R$) of the multiplet R an area law for the $k$-loop with charge $Y_k$,
\eq\ref{wilsonflux}:
\be
 W_k(L)|_{\rm r}=\exp\left[ -(1-\re~ D_R(Y_k)_{\rm r, r})\bar\ell \right].
\ee
The $k$-dependence of the tension due to all members of the multiplet is then proportional to: 
\be
\sigma_k=
\left[d_R-\re\tr \left\{ D_R (Y_k)_R\right\} =
   d_R-\re\tr\left\{\exp(i\pi(Y_k)_R)\right\}\right] l_M n_M.
\label{anyrep}
\ee 
This result is invariant under a gauge rotation,
\be
Y_k\rightarrow \Omega Y_k\Omega^{\dagger}
\ee
as \eq\ref{wilsonflux} suggests. And it reduces to $2k(N-k)$ for the case where R is the adjoint.
The reason is that $(Y_k)_{\rm adjoint}$ has either 0 or $\pm 1$ on the
diagonal as argued already in the previous section.
Hence the formula counts the multiplicity of charged members of the adjoint multiplet.

For the spinor representation one finds from \eq\ref{anyrep} the
result quoted in Ref.~\cite{altes04}:
\be
\sigma_k\sim \left[N-k\cos\left(((N-k)\pi/N\right))-
               (N-k)\cos\left(k\pi/N\right))\right].
\label{spinor}
\ee
Both adjoint and spinor multiplets  are compared 
to the lattice data in the section on data analysis. 

\subsection{Corrections}\label{sec:corrections}

There are two sources of corrections to the Casimir scaling formula.
One is the diluteness, and the other are the effects of Bose-Einstein statistics.
\begin{itemize}
\item The diluteness $\delta=l_M^3n_M=\sigma_1/m_M^2$ is small
($\sim 0.05$, as discussed in section~\ref{subsec:model}) 
but produces corrections. The use
of classical Boltzmann statistics is allowed at large T, since the 
thermal de Broglie wave length $1/T$ is much smaller 
than the inter-particle distance $1/g^(T)T$.
\item As we descend in temperature the diluteness stays constant, since we know from 
the results by Laine and Schroeder~\cite{yorkmikko} that magnetic quantities  
are determined to a very good approximation through all of the plasma phase by 
their value at very large  $T$ in 3d Yang-Mills theory, and the running of the 
coupling due to hard radiative corrections. Below $T_c$ $\sigma_1$ is 
virtually constant~\cite{tepertension05}. Unfortunately the behaviour of the 
magnetic mass is not known in the cold phase, but we know its value at $T=0$
leading to a diluteness $\sim0.09$( see section~\ref{subsec:model}), which suggests that it is 
small at all temperatures.
\item What changes as we go down in $T$ is the ratio of thermal wave length to
inter-particle distance. So Bose-Einstein statistics kicks in at temperatures 
on the order of $4T_c$, where $g^2(T)={\cal O}(1)$. It seems natural that the 
transition is where Bose-Einstein condensation starts.
\end{itemize}
In principle the effects due to the small but non-zero diluteness can be computed.
Comparison to lattice data~\cite{tepertension05} in the deconfined phase shows that they should be small, 
on the order of a few percent at most.  
%%%%%%%%%%%%%%%%%%%%%%%%%%%%%%%%%%%%%%%%%%%%%%%%%%%%%%%%%%%%%%%%%%    

\input{hm.tex}

%%%%%%%%%%%%%%%%%%%%%%%%%%%%%%%%%%%%%%%%%%%
\section{Conclusion\label{sec:conclu}}
%%%%%%%%%%%%%%%%%%%%%%%%%%%%%%%%%%%%%%%%%%%
The picture of hot multi-color QCD considered in this paper
relies on the separation of the hard, soft and ultra-soft 
scales by means of a well-known sequence of two effective theories.
The second of these, which describes the magnetic properties of the 
quark-gluon plasma at energies of order $g^2(T)T$, is the 3D SU($N$) 
gauge theory obtained by dimensionally reducing the original theory. 

In the case of the spatial t'Hooft $k$-loop, 
which records the fluctuations of 
the electric flux of ${\cal N}$-ality $k$ going through it, 
perturbation theory is directly
applicable and becomes ever more accurate at higher temperatures. 
On the other hand, a simple physical picture emerges if the 
electric-flux fluctuations are attributed to gluons passing 
randomly through it. By assuming these quasi-particles to be dilute
and non-interacting, one easily derives an expression whose
parametric dependence on the temperature and the ${\cal N}$-ality
of the loop match the perturbative result.

%Motivated by the idea of duality, 
 The adjoint monopole gas model discussed in this paper
assumes a  similar picture to hold 
for the spatial Wilson $k$-loops: the magnetic-flux 
is attributed to non-Abelian monopoles in the 3D SU($N$) theory. 
The assumption that these monopoles are in the adjoint 
representation directly leads to the prediction of 
Casimir scaling for the ratios of the associated $k$-string tensions.
The $k$-dependence is indeed given
by the multiplicity of charged monopoles with respect to the
charge $Y_k$ of the loop, while the sensitivity to other details 
of the model is reduced in these ratios.
Perturbation theory is not applicable in this sector, but
non-perturbative lattice Monte-Carlo calculations of the
$k$-string tension ratios, although numerically challenging, 
are in principle straightforward.

Previous simulations~\cite{lucini} for $N=4,~6$, as well as the $N=8$ data
presented in this paper confirm the Casimir scaling property
of $k$-string tension at the few percent level. Care must be taken in these
calculations that the strings are long enough, $\sigma L^2\gg N$, 
for the quantum corrections to the string energy to be subleading
with respect to the weak binding energy of the $k$ fundamental strings.
The energy of the string is then large and a multi-level algorithm~\cite{2leva}
proved useful in this situation to reduce the variance 
on the correlator from which this energy is extracted.
While the Casimir scaling prediction lies slightly above the 3-loop
expression for the t'Hooft loop, it is slightly lower than
the lattice results for spatial Wilson loops: the (small) corrections
to Casimir scaling in the magnetic sector 
seem to have the opposite sign with respect to the electric sector.

The fundamental assumptions of the model can be further tested.
The behaviour of $k$-loops, as we argued, is the simplest observable
to consider. It can be measured for other gauge groups, as long as 
the center is $Z(N),~N \geq 4$;
indeed the adjoint monopoles may be formed in \emph{any} 
non-Abelian gauge theory. 
The classical groups have as candidates, apart from the SU($N$) groups, 
the Spin$(4p+2)$ groups with center group $Z(4)$.
One may also introduce an adjoint Higgs field which acquires a VEV. 
Depending on the symmetry breaking pattern, 
some monopoles will become heavy, and the $k$-ratios will change
in a predictable way~\cite{rajantiealtes}.  

We also discussed the behaviour of the 't Hooft loops for different ${\cal N}$-alities
at finite temperature. Here the same Casimir scaling is observed in the lattice
data~\cite{lucinidefor04,teper05} as predicted by perturbation theory for high $T$
(and by the quasi-particle picture of gluons). 
Surprisingly, the scaling continues to hold down to practically $T_c$.
And the same is true for the magnetic k-loops~\cite{yorkmikko,tepertension05}: 
Casimir scaling stays valid down to $\sim T_c$.

In conclusion: all available  data on electric and magnetic k-loops for $N\ge 4$
are consistent with a screened electric and magnetic quasi-particle model 
throughout all of the plasma phase. Whether the same  
is true for QCD ($N=3$) remains to be tested.

\paragraph{Acknowledgements\\}
We acknowledge discussions with Pierre van Baal,
Sander Bais, Nick Dorey, Philippe de Forcrand, Pierre Giovannangeli, Prem Kumar, 
Mikko Laine, Nick Manton, Hugh Osborne, Owe Philipsen, Martin Schvellinger, 
Jan  Smit, and Mike Teper. Finally to Tony Kennedy, Tony Gonzalez-Arroyo and 
participants of the lattice Meeting (March 2005 at KITP Santa Barbara).

The lattice simulations were performed on the PC cluster of the Rudolf Peierls Centre
for Theoretical Physics at Oxford University in the year 2004. 
The machine was partly funded by EPSRC and PPARC grants. 
%
%%%%%%%%%%%%%%%%%%%%%%%%%%%%%%%%%%%%%%%%%%%%%%%%%%%%%%%%%%%%%%%%%%%%%%%%%%%%

%%%%%%%%%%%%%%%%%%%%%%%%%%%%%%%%%%%%%%%%%%%%%%%%%%%%%%%%%%%%%%%%%%%%%%%%%%
\appendix

\section*{Appendix A}

In this appendix we briefly indicate the group theory needed to get from a given
Young tableau (defining the irreducible representation $R$) the corresponding 
highest weight and the value of the quadratic Casimir. In what follows we suppose
a representation to be irreducible without mentioning so.   
We choose the $Y_k$ as follows:
\be
Y_k = {1 \over N} ~{\rm diag}(\underbrace{k,k,\dots,k}_{N-k~{\rm times}},
       \underbrace{k-N,k-N,\dots,k-N}_{k~{\rm times}}). 
\ee

Let the Young tableau have $n_1$ boxes in the first row, $n_2$ in the second row, 
etc\dots Then one can define the non-negative numbers $w_l=n_l-n_{l+1}$.
Now the highest weight matrix for the Young tableau is defined through the $Y_k$ matrices: 
\begin{equation}
H_R=\sum_{l=1}^{N-1}w_lY_l.
\label{highweight}
\end{equation}
For example, for the totally antisymmetric tableau of k boxes in one column we 
have $H_R=Y_k$. For the totally symmetric tableau with all k boxes in one row $H_R=kL_1$.
Note that the stability group of $Y_k$ (the subgroup of SU($N$) matrices 
commuting with $Y_k$) is ${\rm SU}(k)\times {\rm SU}(N-k)\times {\rm U}(1)$.
So the totally antisymmetric representation with k squares has a highest weight
with this stability group. All other representations with k squares have different stability groups.

We define one more diagonal $N\times N$ matrix by:
 \begin{equation}
2Y\equiv 2\sum_{l=1}^{N-1}Y_l=\diag (N-1,N-3,....,-N+1).
\end{equation}
The quadratic Casimir operator $C_2(N,k,\{w_l\})\equiv C_2(R)$ is defined
by summing the square of all generators $T_a$ in the representation $R$. 
The result is $\sum T_a^2=C_2(R) 1_{R}$, where $1_{R}$ is the unit matrix in 
$R$ and $C_2(R)$ is a c-number (normalization is 
$[T_a,T_b]=if_{abc}T_c, f_{abc}f_{bcd}=N\delta_{ad}$).
Then  the quadratic Casimir equals:
\begin{equation} 
C_2(R)={1\over 2}(\tr\{H_R^2\}+2\tr\{YH_R\}).
\end{equation}
The quadratic Casimir for the fundamental representation is $C_F={(N^2-1)\over{2N}}$.
The Casimir for the antisymmetric representation is then
\begin{equation} 
C_2(R=AS)=C_F{k(N-k)\over{(N-1)}},
\end{equation}
and for the symmetric representation it is:
\begin{equation} 
C_2(R=SS)=C_F{k(N+k)\over{(N+1)}}.
\end{equation}
To derive these relations one needs the inner product of two $Y$ matrices:
\begin{equation}
\tr\{Y_kY_l\}={1\over N}(\min(k,l)N-kl).
\end{equation}
One can show that for fixed $k\le N$ the antisymmetric Casimir is the 
minimal one.

Finally we give the relation between the $Y_k$ matrices and the Chevalley basis
$H_{k,k+1}=\diag (0,0,0,....1,-1,0,...0)$, $Y_{N,N+1}\equiv Y_{N,1}$.
We introduce the matrices $y_k$, with $y_1\equiv NY_1$. $y_2$ follows from $y_1$
by a cyclic permutation of the diagonal elements: the first becomes the second 
and so forth. $y_3$ follows from $y_2$ the same way. We keep doing this until 
we have reached $y_N$, $y_{N+1}=y_1$. The sum of all the $y_k$ vanishes.
The $Y_k$ are related to the $y_k$ by:
\be
NY_k=\sum_{l=1}^ky_l.
\label{sumy}
\ee
Then:
\ba
{1\over 2}y_1&=&(N-1)H_{12}+(N-2)H_{23}+(N-3)H_{34}+.....+1.H_{N-1N}+0.H_{N1}\nonumber\\
{1\over 2}y_2&=&0.H_{12}~~~+(N-1)H_{23}+(N-2)H_{34}+.....+2.H{N-1N}+1.H_{N1}\nonumber\\
{1\over 2}y_3&=&1.H_{12}~~~~~~~~~~~+0.H_{23}+(N-1)H_{34}+.....+3.H{N-1N}+2.H_{N1}\nonumber\\
\nonumber
\ea
The general term is:
\be
{1\over 2}y_k=(k-2)H_{12}+(k-3)H_{23}+...+0.H_{kk-1}+(N-1)H_{kk}+....+kH_{N-1N}+(k-1)H_{N1}.\\
\ee
Note the diagonal translation invariance of the coefficient matrix $M$ relating 
the $N$ $y_k$ to the $N$ $H_{l,l+1}$.

Evaluate the first diagonal element  of the r.h.s. in \eq\ref{sumy}, 
using the matrix $M$ above. It equals, because of the relative sign
in the non-zero elements of $H_{12}$ and $H_{N1}$, the difference of the first 
column and the last column of $M$, up and including the first $k$ rows of $M$. 
Because of the translation invariance only
the difference $M_{11}-M_{kN}=N-k$ survives. The second diagonal element equals
$M_{k1}-M_{12}=N-k$. This goes on till we reach the coefficient $M_{kk-1}=0$.
Then there is a jump to $M_{k,k+1}=N-1$, and the diagonal elements become $-k$.
So we reproduced the matrix $NY_k$ in  \eq\ref{sumy}.
The charges $NY_k$ lie on the root lattice spanned by the $H_{ll+1}$.

%\section*{Appendix B}
%In this appendix we review the Stokes formula~\cite{diakonov} for the Wilson loop in the representation $R$. Let $\Omega$ be any gauge transformation that is periodic on the loop. Then, with $\vec\nabla\Omega^{\dagger} =(\vec\partial-ig\vec A)\Omega^{\dagger}$ and $\vec\nabla\Omega=\vec\partial\Omega+ig\Omega \vec A$:
%\begin{equation}
%W_R(L)=\int D\Omega\exp{\{ig\int d\vec S.\tr[H_R\big(\Omega\vec B\Omega^{\dagger}-{1\over {ig}}\vec\nabla\Omega\times \vec\nabla\Omega^{\dagger}\big)]\}}.
%\label{diakonov}
%\end{equation}

%If $R$ is the antisymmetric representation we have from the formulae for the highest weight in Appendix A the equation in the text, \eq\ref{wilsonflux}.
%Note that this representation is  one with stability group  $SU(k)\times SU(N-k)\times U(1)$. The same stability group apears in all representations with rectangular Young diagrams with k rows and l columns. Their weights are according to \eq\ref{highweight} $lY_k$.

 %The integration over the regular gauges has been dropped in the main text because physical states do not feel them. Physical states include monopole configurations, which is why the second term in \eq\ref{diakonov}
%is discarded in eq\ref{wilsonflux}.

\section*{Appendix B}
The derivation of \eq\ref{diakonov} is based  on simple properties of the 
quantum-mechanical SU($N$) rotator in an external field. We reproduce it here in a
form that should render its origins clear.

We are interested in the Wilson line between two points $x(s_1)$ and $x(s_2)$, 
the line between the two points being parametrized by $s$. The line 
is the ordered product (from right to left) in some irreducible representation
$R_0$ with highest weight $H^0$ of unitary matrices with dimension $d_{R_0}$:
\begin{equation}
W(s_f,s_i)=P ~ \exp \left[ ig\int_{s_i}^{s_f} A_sds \right].
\end{equation}
Here $A_s={d\vec x\over {ds}}.\vec A$ the projection of $\vec A$ on the
line. 

The Wilson loop is covariantly constant along the curve L, 
$\partial_s W^{\dagger}(s,s_i)-igW^{\dagger}A_s=0$, so
 along the loop one   can write the vector potential as a pure gauge
\begin{equation}
 A_s={-1\over{ig}}U\partial_sU^{\dagger}
\label{aspure}
\end{equation}
with $U=W(s,s_i)U_i$ with $U_i$ an arbitrary SU($N$) matrix.  
So the loop becomes in the representation $R_0$ the unitary matrix  :
\begin{equation}
D^{R_0}(U_fU_i^{\dagger}).
\label{puregauge}
\end{equation}
We close the Wilson loop ( so $s_f$ and $s_i$ represent the same point) 
and take the trace of \eq\ref{puregauge}. The main result of this appendix is: 
\begin{itemize}
\item The sum over all irreducible representations of the normalized trace of the Wilson loop is the group average 
 of the   propagator of the SU($N$) rotator with 
Hamiltonian ${\cal{H}}$ in the external field $H^0$, \eq\ref{propagator}.
\item Then a special limit of the path integral version
of this propagator, \eq\ref{pathform}, is nothing but \eq\ref{diakonov}.
\end{itemize}

To define the Hamiltonian we start with the generators $l_a$ of the group SU($N$). They
act as left multiplication on the Hilbert space defined on the group manifold, with kets $|U\rangle$,
$U$ an SU($N$) element. The generators obey the commutation relations
\begin{equation}
[l_a,l_b]=if_{abc}l_c, ~~\mbox{with}~~f_{abc}f_{a'bc}=N\delta_{aa'}.
\label{groupgenerators}
\end{equation}

The $N-1$ diagonal generators $l_d$ are denoted by the suffix $d$, whereas 
a general generator is denoted by indices $a$. 

Likewise we can define the generators $r_a$ of right multiplication.

The state space of kets $|\Omega>$ is has an complete orthogonal basis consisting
of all the irreducible representations $D^R(\Omega)_{l,r}$, and the orthogonality relations read:
\begin{equation}
\int d\Omega ~ D^R_{k,l}(\Omega)D^{R^{\prime}}_{m,n}(\Omega^{\dagger})=
{1\over{d_{R_0}}}\delta_{R,R^{\prime}}\delta_{k,n}\delta_{l,m}.
\label{ortho}
\end{equation}
The dimension of the representation R equals $\tr\{D^R({\bf{ 1}})\}=d_R$.
We define kets and bras $|R;l,r \>$ with the property:
\begin{equation}
\< \Omega | R;l,r \> = D^R(\Omega)_{l,r}.    \label{rbasis}
\end{equation}
From \eq\ref{ortho} one sees that the norm of such kets is $d_R^{-1}$.

The Hamiltonian is defined in terms of the left multiplication generators as:
\begin{equation}
{\cal{H}}={1\over{2I}}\Big(\sum_{a}l_a^2-\sum_dl_d^2+\sum_d(l_d-{1\over 2}H^0_d)^2\Big).
\label{hamiltonian}
\end{equation}
The last term represents the coupling of the rotator to the highest weight 
$H^0$ written in component form $H_0=\sum_dH^0_d{\lambda_d\over 2}$, which gives
 a Zeeman effect for the energy levels in any representation R.

Write $\sum_al_a^2=C_2(R)$ for the value of the quadratic Casimir in R and denote
$l$ for the $N-1$ diagonal quantum numbers $l_d$ in $D(\Omega)_{l,r}$.  Similarly
the $N-1$ diagonal right multiplication generators have quantum numbers, denoted by $r$.
Then one has for the eigenvalues $E^0$  of this Hamiltonian:
\begin{equation}
E^0(R,l)={1\over{2I}}\big(C_2(R)-\sum_dl_d^2\big)+{1\over{2I}}\sum_d(l_d-{1\over 2}H^0_d)^2.
\label{energy}
\end{equation}
Let us find the corresponding  Lagrangian $L$ . Define the angular velocities 
from the special unitary matrices S:
\begin{equation}
V_a= i ~ \tr\{S\partial_s S^{\dagger}\lambda_a\}.   \label{ang}
\end{equation}
Then:
\begin{equation}
L={I\over 2}\sum_aV_a^2+{1\over 2}\sum_d V_d H^0_d.
\label{lagrange}
\end{equation}
This Lagrangian gives the Hamiltonian in terms of the canonical momenta 
$J_a={\partial L\over{\partial {V_a}}}$. There is a subtlety: upon quantization 
we have the generators of left and right SU($N$) rotations, related by the adjoint representation.
The question is then to which the J's do correspond. This question is only relevant
 for the linear term,since the quadratic term is invariant under the adjoint. 
A short calculation shows that the left generators correspond to the J's. 

Let us now prove the following relation between the characters of the group and 
the integrated propagator of the SU($N$) symmetric top:
\begin{eqnarray}
&{}&\int d\Omega ~ \langle \Omega U_f|\exp\left\{i(s_f-s_i){\cal{H}}\right\}|\Omega U_i\rangle\nonumber\\
 &=&\sum_{R,l} \tr\{ D^{R} (U_f U_i^{\dagger})\} ~ \exp\left[i(s_f-s_i)E^0(R,l)\right].
\label{propagator}
\end{eqnarray}

On the r.h.s. we integrate over SU($N$) matrices $\Omega$, with the measure normalized to 1.

To prove this, insert the set of intermediate states $|R;l,r>$ from \eq\ref{rbasis} into the l.h.s. :
\begin{eqnarray}
 &{}&\int d\Omega ~ \langle\Omega  U_f| \exp\left\{ i(s_f-s_i){\cal{H}}\right\}|\Omega U_i\rangle
    \nonumber \\ 
 &=&\int d\Omega \sum_{R, l,r}d_R ~ \langle\Omega U_f|R;l,r\rangle ~ \exp \left[ i(s_f-s_i)E^0(R,l)\right]
~ \langle R; l, r|\Omega U_i\rangle.
\label{intermediate}
\end{eqnarray}
The right index $r$ can be summed over, and using  \eq\ref{rbasis} one finds:
\begin{equation}
\sum_r\langle\Omega U_f|R;l,r\rangle \langle R; l, r|\Omega U_i\rangle=
 D^R(\Omega U_f(\Omega U_i)^{\dagger})_{l,l}.
\end{equation}
Integration over $\Omega$ gives, for fixed $l$, using \eq\ref{ortho}:
\begin{equation}
\int d\Omega ~ D^R(\Omega U_f(\Omega U_i)^{\dagger})_{l,l}={1\over {d_R}}\tr\{D^R(U_fU_i^{\dagger})\}
\end{equation}
Plug this result into the r.h.s. of \eq\ref{intermediate}, and we have the 
main result \eq\ref{propagator}. 
 
Now we want to project out from the main result the representation $R_0$. 
This is done by letting  $I\rightarrow 0$ in the energy exponent 
and determining the minimum of $E^0(R,m)$. 
One such minimum is realized by ${1\over 2}H^0_d=m_d$ and $C_2(R)=C_2(R_0)$.

So the sum over all irreducible representations R in \eq\ref{propagator} 
reduces in this limit to the representation $R_0$ singled out by the 
its corresponding highest weight $H^0$.

Strictly speaking there is only one minimum   when $H_0$ is the weight of the 
fundamental representation, or of any of the representations corresponding to a 
fully antisymmetric Young tableau. The reason is that for fixed ${\cal N}$-ality $C_2(R)$
takes its minimum value for R being fully antisymmetric.

For higher representations, like e.g. in SU(2) with weight $j_0$( the number of 
boxes in the Young tableau), there is the representation $R=j_0-1$, with $m=j_0-1$,
that minimizes $E^0$ as well.  This is why in the original work~\cite{diakonov} 
the \emph{asymmetric} top was taken, to provide an independent inertia in front 
of the Zeeman splitting in \eq\ref{energy} and taking them independently to zero. 
So from now on the irrep $R_0$ stands for one of the totally antisymmetric representations.

Of course the result contains the rapidly varying  phase factor 
 $$\exp \left[ i(s_f-s_i) {E^0(R_0,H_0)\over{2I}} \right] =
\exp \left[ i(s_f-s_i)\left(C_2(R_0)-{1\over 2}\tr\{H_0^2\}\right)/2I \right] \equiv F(R_0).$$
According to appendix A the value in this minimum is 
$C_2(R_0)-{1\over 2}\tr\{H_0^2\}=\sum_{l=1}^{N-1}w_l(N-l)lN$ 
with $w_l$ the weights defined by the Young diagram of $R_0$.
This same factor $F(R_0)$ appears also in front of the path integral
transcription of the matrix element:  
\begin{equation}
\langle \Omega U_f| \exp \left\{i(s_f-s_i){\cal{H}}\right\} |\Omega U_i\rangle=
F(R_0)\int_{\Omega U_f}^{\Omega U_i} DS(s)~\exp \left[-i\int ds L \right]
\label{pathform}
\end{equation}
with $L$ as in \eq\ref{lagrange}.
This formula, together with the fact that the left generators 
in the matrix element are correctly  represented by the path integral can be proved straightforwardly.
We introduce the periodic fluctuation variable $\Omega(s)$ with $\Omega(s_{i,f})
=\Omega$ and transform the fluctuation variable $S$  in \eq\ref{pathform}:
\begin{equation}
S(s)=\Omega(s)U(s){} \qquad \mbox{with} \quad D\Omega=DS.
\label{substitution}
\end{equation}
The path integral now becomes:
\begin{equation}
\int_{\Omega}^{\Omega} D\Omega(s) ~ \exp\left[ -i\int ds L\right]
\label{pathformbis}
\end{equation}
with $L$ as in \eq\ref{lagrange}, 
but with the substitution \eq\ref{substitution} in the angular velocities $V_a$ in (\ref{ang}):
\begin{eqnarray}
 V_a&=&\tr\{\Omega(s)U(s)\partial_s(\Omega(s)U(s))^{\dagger}\lambda_a\} \nonumber\\
&=&\tr\left\{\left(\Omega(s)(U(s)\partial_sU(s)^{\dagger})\Omega(s)^{\dagger}+
\Omega(s)\partial_s\Omega(s)^{\dagger}\right)\lambda_a \right\}.
\label{angular}
\end{eqnarray}
Finally use \eq\ref{aspure} to write the angular velocity as the gauge transformed potential $A_s$:
\begin{equation}
V_a=\tr\left\{\left(\Omega(s)(-igA_s)\Omega(s)^{\dagger}+\Omega(s)\partial_s
\Omega(s)^{\dagger}\right)\lambda_a\right\}=
\tr\left\{ \left(\Omega(s)\nabla_s\Omega(s)^{\dagger}\right)\lambda_a \right\}.
\end{equation}
Remember $\nabla_s\Omega(s)^{\dagger}={d\vec x\over{ds}}.(\vec\partial-ig\vec A)\Omega(s)^{\dagger}$.

The final form of the   Wilson loop in the irrep $R_0$ and with highest weight 
$H_0$  follows from  \eq\ref{lagrange}, \ref{propagator} and  \ref{pathform}:
\begin{equation}
W_{R_0} = \lim_{I\rightarrow 0}
       \int D\Omega ~\exp \left[-i{I\over 2}\oint ds\sum_aV_a^2-\oint ds ~
          \tr\left\{H_0(\Omega\vec\nabla \cdot {d\vec x\over{ds}}\Omega^{\dagger})\right\}\right].
\end{equation}
The line integral on the r.h.s. can be easily transformed 
into a surface integral, because it now involves the gauge 
transform average of  potential projected on the highest weight, i.e. an Abelian potential.   
The surface S bounded by the loop $L$ is covered by a set of nested loops 
$L(r)$ with $L(r=0)=L$ and $L(r=1)$ shrunk to a point. 
Then, in an obvious notation:
\begin{eqnarray}
&{}&-\oint_L ~ ds ~ \tr\{ H_0(\Omega \nabla_s\Omega^{\dagger})\}\nonumber \\ 
&=&\int_0^1 dr ~ \partial_r\oint ds~\tr\{H_0\Omega \nabla_s\Omega^{\dagger}\}\\ \nonumber
&=&\int_0^1 dr\oint ds ~ \tr\left\{ H_0\left(\epsilon_{rs}\nabla_r\Omega \nabla_s
 \Omega^{\dagger})+ \epsilon_{rs}\Omega\nabla_r \nabla_s\Omega^{\dagger}\right)\right\}
\end{eqnarray}
which gives then \eq\ref{diakonov}, using the commutator $[\nabla_r,\nabla_s]=-igF_{r,s}$.
The gauge transform in the last expression is now extended to all of the surface
S and beyond. This surface is of  course arbitrary, apart from its boundary.

%%%%%%%%%%%%%%%%%%%%%%%%%%%%%%%%%%%%%%%%%%%%%%%%%%%%%%%%%%%%%%%%%%%%%%%%%%%
%\clearpage
\section*{Appendix C}
\input{appendix_C.tex}
%%%%%%%%%%%%%%%%%%%%%%%%%%%%%%%%%%%%%%%%%%%%%%%%%%%%%%%%%%%%%%%%%%%%%%%%%%%
\input{tab.tex}
\input{fig.tex}
\end{document}

%% file: hm.tex
%%%%%%%%%%%%%%%%%
\section{On strings in higher representations  and ${1\over N}$ corrections
\la{sec:string_high_rep}}
%%%%%%%%%%%%%%%
We now leave the discussion of the adjoint-monopole-gas model and 
discuss the properties of $k$-strings from the point of view of the 
large-$N$ expansion.

Standard arguments on large $N$ SU($N$) gauge theory~\cite{hooft,witten_baryons}, 
based on the planarity of Feynman diagrams and the (assumed) confinement of color, 
imply that gauge invariant states have masses of order $N^0+N^{-2}$, with
a width of order $N^{-2}$. Also, no bound state of color singlet constituents
survives the large \n limit: the theory is expected to be a theory of 
free `hadrons'.

It is interesting to consider, at large but finite number of colors, 
precisely those states whose wavefunction contains a 
significant component which is a direct product of color singlet pieces.
Phenomenology provides a number of potential examples.
A classic example would be the deuteron, a very loosely bound state
lying only a few MeV under the nucleon-nucleon threshold. Another interesting
though less firmly established case is the $f_0(980)$ meson, 
whose wavefunction has been discussed in terms of a mixture of a kaon-kaon molecule
and a four-quark state~(\cite{Close:2002zu} and ref. therein). 
Again, the state is only a few MeV under the two-kaon threshold. 

At a more theoretical level, there are examples in the pure SU($N$) gauge theory
in three and four dimensions.
Consider the theory defined on a finite (but large: $L\gg 1/T_c$) hypertorus. 
In addition to glueballs, the spectrum contains `torelon' states 
(whose mass we denote by $m_k(L)$)
which transform non-trivially under the centre symmetry $Z_N$. 
The sectors of different $\cal N$-ality are protected by this global symmetry.
Thus for $N\geq4$, one may ask whether two fundamental torelons can form
a bound state lying under the threshold $2m_{k=1}(L)$. That this is indeed
the case was first numerically demonstrated from first principles in
the work~\cite{lucini}. Further, at large $L$
the states are string-like and one can ask what the ratios of their string
tensions are (we may use the fundamental `$k=1$' string tension as the reference). 
An alternative formulation of the problem would consider the strings to
be open and attached to static sources in the appropriate 
representation~\cite{Deldar:1999vi}. 

For simplicity, we now focus on the $k=2$ sector;
for $N \geq 4$, the screening of the string is forbidden by the centre symmetry.
In our view, the first question to settle in the context of the large-$N$ 
expansion is, `What is the $1/N$ power of the leading correction to
the planar limit result $\sigma_2=2\sigma_1$?'. 
Since $m_2(L)$ lies under the threshold $2m_{k=1}(L)$ at all $L$~\cite{lucini}, 
the question arises whether one should think of the $k=2$ torelon as a 
weakly bound state of two $k=1$ torelons,
or if the colour structure gets completely rearranged 
into a single `unfactorisable' color singlet piece.
In the nucleon-nucleon system, the analogous question is whether the deuteron is
primarily a bound state of two nucleons, or a 6-quark state.
In the first case, considered in~\cite{armoni},
the long-distance attractive force between the two $k=1$ strings
will be driven by the exchange of the lightest
($0^{++}$) glueball, while the short distance force is essentially given by 
two-gluon exchange. Both effects are indeed~\cite{armoni} 
suppressed by $1/N^2$ with respect to the free propagation of two $k=1$ strings. 
Regarding the second configuration, the simplest classical string configuration
is that of a single string winding twice around a cycle of the hypertorus. 
At large \n, the energy of such a configuration is expected to approach
 threshold from below at a $1/N^2$ rate.

At finite \n and asymptotically large $L$ however, 
we are in presence of two almost degenerate configurations lying near threshold. 
It is therefore imperative to consider 
the mixing effects between these two configurations.
To keep the discussion as simple as possible, we may keep the transverse spatial
dimensions $L_\perp$ finite, so as to separate two-torelon `scattering states' from
the weakly bound states we discuss by a finite, fixed gap. As \n is increased, 
this transverse volume can be increased as well without affecting the validity of
our treatment of the $k=2$ sector as a 2-level system.

We suppose, following~\cite{finivol}, that the Hamiltonian of the SU($N$)
gauge theory can be expanded in inverse powers of $1/N$:
\be
H(L,N)=\sum_{k=0}^\infty \frac{H_k(L)}{N^k}.
\ee
The existence of the t'Hooft limit  implies that $H_o(L)$ has the same eigenvalues as
the Hamiltonian of the SU($\infty$) theory in the same spatial volume. 
We consider now the $k=2$ flux tubes
winding around a cycle of the torus as a quantum mechanical two-state system, 
as was done in~\cite{finivol} for the case of the scalar glueball -- 
adjoint Polyakov loop system in intermediate volume.
Consider on the one hand the state made of two $k=1$ non-interacting
closed fundamental strings, and on the other a single fundamental
string with winding number 2; in this basis 
$H_o$ reads $H_o=2m_{k=1}{\bf I}_{2\times2}$. 

The `perturbation' describes the deviations from the planar limit. On the diagonal,
the corrections are ${\cal O}(1/N^2)$. Indeed, the attractive potential between 
two fundamental torelons is suppressed by the product of two 3-vertices 
each of which carries a $1/N$ factor. On the other hand, the amplitude of the 
transition from one of our basis states to the other only contains one such vertex, 
and therefore the off-diagonal element of our $2\times2$ hamiltonian is 
${\cal O}(1/N)$. 
The perturbation hamiltonian in our basis reads:
\be
% \Delta H= \left(\begin{array}{c@{~}c} \frac{-\bar h_1}{N^2}&\frac{\bar h}{N}\\
% \frac{\bar h}{N} & \frac{-\bar h_2}{N^2} \end{array}\right)
\Delta H= \left(\begin{array}{c@{~}c} -\bar h_1/N^2 & \bar h/N\\
\bar h/N & -\bar h_2/N^2 \end{array}\right)
\la{eq:DH}
\ee
with $\bar h_1$, $\bar h_2$ and $\bar h$ of order $N^0$. It is clear that 
to leading order in $1/N$, the resulting
energy eigenstates are now the symmetric and anti-symmetric linear combinations
of our basis states. The associated energies are 
$E_A=2m_{k=1}-\frac{\bar h}{N}+{\cal O}(1/N^2)$ and 
$E_S=2m_{k=1}+\frac{\bar h}{N}+{\cal O}(1/N^2)$. We thus reached the perhaps 
surprising conclusion that the corrections to the mass of the lightest
$k=2$ string are of order $1/N$. There is one state below threshold and one above,
situated symmetrically about the threshold energy, up to $O(1/N)$ corrections.
We note that $\bar h$ (as well as the $\bar h_i$) 
is expected to grow proportionally to $L$ at large $L$,
since the breaking of the string can occur at any point along the string,
so that the ratio in the torelon masses directly translates into the ratio of the
string tensions.

A caveat particularly relevant to Monte-Carlo simulations
is that in all the considerations above 
we have supposed the strings to be long enough 
to be able to identify the ratios of string tensions ratios
with the ratio of torelon masses. Since the $1/L$ string correction~\cite{luscher}
lowers the energy of the string, it induces a repulsive force between
two fundamental torelons at finite $L$~\cite{finivol}. 
Since the binding energy of the strings
reduces at large \n, $L$ must be increased so that the condition 
\be
\sigma_1 L^2 \gg N   \la{eq:long_s_cond}
\ee
is satisfied to ensure that the ratio of torelon loops yields the correct
$N$-dependence of the string tension ratios. 
If the large $N$ limit is taken at finite $L$, the ratio of 
$k=2$ to $k=1$ torelon masses will approach 2 with $1/N^2$ corrections, because 
a mixing amplitude only affects the energy spectrum at leading order if the 
`unperturbed' states are degenerate (to that order). And indeed, the 
two classical-string configurations we took as unperturbed states have different
$1/L$ corrections: if $ \gamma_1$ is the L\"uscher coefficient of the fundamental
string, the direct-product configuration of two $k=1$ strings
has a $2\gamma_1/L$ correction, while the fundamental string with winding number 2
admits a $\gamma_1/2L$ string correction.
As  all lattice simulations so far~\cite{lucini,debbio,lucini04}, 
ours are done in the regime $N < \sigma_1 L^2 < N^2$.
The  second inequality implies that the energy gap 
$(\bar h_1-\bar h_2)/N^2$ is parametrically smaller than the string 
vibrational excitations.

\eq\ref{eq:long_s_cond} however also implies  that the vibrational 
excitations of the strings are separated by $4\pi/L$ gaps which are parametrically
much smaller than the mixing energy $\bar h/N$. 
We  have thus neglected the matrix elements of the Hamiltonian $H_1$ 
between the two states that we focused on and 
the vibrationnally excited states, since we reduced the 
diagonalisation problem from the full Hilbert space to the space spanned 
by these two states. Although the neglected matrix elements 
can modify the splitting pattern around the threshold,  the 
mixing between the string ground states will be enhanced relatively to the 
other mixings if the condition $\sigma_1 L^2 \ll N^2$ holds. In any case,
the parametric size of these matrix elements is 
$\sigma L/N$ and so we must expect the corrections to the $k=2$ string tension
to be of that order.

In summary, the predictions of the two-state mixing model are:
\begin{enumerate}
\item the energy eigenstates are the anti-symmetric and the symmetric linear 
combinations of the direct-product configuration of two $k=1$ strings
and the fundamental string with winding number 2.
\item they are split symmetrically around the threshold energy $2m_1$ (up to 
$O(1/N)$ corrections).
\item the splitting energy itself is of order $1/N$.
\end{enumerate}
%%%%%%%%%%%%%%%%%%%%%%%%%%%%%%%%%
\subsection{Static potentials}
%%%%%%%%%%%%%%%%%%%%%%%%%%%%%%%%%
If one considers open strings attached to static `quarks', 
the argument takes a slightly different form. The relevant quantities 
here are the static potentials between colour sources
in \emph{irreducible} representations of SU($N$).

The k=1 string binds a quark and a distant antiquark together. 
Similarly  the k=2 configuration can be viewed as two (weakly interacting) strings 
each joining one of the quarks to one of the antiquarks. 
If we number the quarks by 1 and 2, and the antiquarks by $\bar 1 $ and $ \bar 2$,
then there are two classical string configurations which are exactly degenerate:
the configuration where $1$ is attached by a string to $\bar 1$ and
$2$ is attached to $\bar 2$, and the other where $1$ is attached to $\bar 2$ 
and $2$ to $\bar 1$. However, 
the interaction between the strings can take one configuration into the other. 
Therefore a splitting occurs between the symmetric and anti-symmetric 
linear combinations, corresponding to the static potential  
splitting between the $k=2$ symmetric
and anti-symmetric irreducible representations of SU($N$). 
There is however general agreement that screening of 
the static sources through virtual gluons implies that the string tension 
obtained in either representation at large enough separations 
is the same; although it can be difficult
to demonstrate this in Monte-Carlo simulations.

%%%%%%%%%%%%%%%%%%%%%%%%%%%%%%%%%%%%%%%%%%%%%%%%%%%%%%%%%%%%%%%%%
\subsection{A caveat on the implications of factorisation at large $N$}
%%%%%%%%%%%%%%%%%%%%%%%%%%%%%%%%%%%%%%%%%%%%%%%%%%%%%%%%%%%%%%%%%
The standard way to extract the static potential for
fundamental charges, namely by measuring the expectation value 
of a rectangular Wilson loop of size $R\times T$, $T\gg R$, can be generalised
to extract it for any representation~\cite{Deldar:1999vi}. In particular, the simplest way
to obtain a representation of ${\cal N}$-ality $k=2$ is to take the real part of
the square of $W(R,T)$, the trace of the fundamental Wilson loop. 
At finite $R,~T$, the factorisation property of gauge invariant operators
(see for instance \cite{Das:1984nb})
then implies that the expectation value of this operator is given by 
\be
\< W(R,T)^2 \>=\< W(R,T) \>^2~(1+{\cal O}(1/N^2)).\quad \mathrm {(factorisation)}
\ee
On the other hand, if we consider small separations $R$, asymptotic freedom
implies that the short-distance potential in an irreducible representation 
${\cal R}$ is given by $C_{\cal R}\alpha_s/R$.
The symmetric $k=2$ representation has $C_{\cal R}=C_S=2(N+2)C_F/(N+1)$, while 
the $k=2$ anti-symmetric has $C_A=2(N-2)C_F/(N-1)$. In particular, for the 
fundamental representation, it is 
\be
W(R,T)=\exp{\left(-\frac{\bar \alpha T}{2R}\right)} ~+~ {\cal O}(1/N^2),
\qquad R\ll \sigma^{-1/2}.
\ee
The operator $W^2(R,T)$ belongs to a representation that can be reduced 
into the symmetric and anti-symmetric. Therefore, if we take the 
$T\rightarrow\infty$ limit, the potential energy of 
the anti-symmetric representation dominates the expectation value of $W^2(R,T)$:
\be
\lim_{T\rightarrow\infty} \<W^2(R,T)\>
 \propto e^{-C_A\alpha_s CT/R}=\<W(R,T)\>^2 ~e^{\frac{\bar\alpha T}{NR}}
~(1+ {\cal O}(1/N^2)),  ~~ R\ll \sigma^{-1/2}.
\ee
Thus tree-level perturbation
theory contradicts the large-$N$ counting rules concerning the leading corrections
to factorisation. The origin of the paradox lies in the straightforward
$T\rightarrow\infty$ limit necessary to filter out the ground state.
If the contribution from the symmetric representation is kept, 
the large $N$ limit of the small-$R$, large-$T$ Wilson loop
$\<W^2(R,T)\>$ is given by
\be
\lim_{T\rightarrow\infty} \<W^2(R,T)\>  \propto 
\<W(R,T)\>^2 ~\cosh\left(\frac{\bar\alpha T}{NR}\right)
~(1+ {\cal O}(1/N^2)),
\ee
which, at fixed $T$, has $1/N^2$ corrections to the planar limit result.

What have we learnt? The large-$N$ factorisation property 
does not necessarily imply that the lowest energy state of a `meson' 
made of a static colour source in a certain representation and its anti-source
has ${\cal O}(1/N^2)$ corrections, 
\emph{because other representations of same ${\cal N}$-ality become degenerate
with it in the $N\rightarrow\infty$ limit}. Since string tensions are extracted
from the lowest energy at large $R$, the same caveat applies to them.
%%%%%%%%%%%%%%%%%%%%%%%%%%%%%%%%%%%%%%%%%%%%%%%%%%%
\subsection{Strings in open and closed form\la{sec:open-closed}}
%%%%%%%%%%%%%%%%%%%%%%%%%%%%%%%%%%%%%%%%%%%%%%%%%%%%%%
We finish with a remark on the relation between different representations
of same ${\cal N}$-ality and excited states in the open and the closed string sectors.
To that end it is useful to consider the correlator of Polyakov loops of length $L$
$\< P_{\cal R}(0) P_{{\bar{\cal R}}'}(\vec x)\>$. This expectation value is interpreted 
(from the point of view of the transfer matrix along the dimension of size $L$) 
as the free energy of the system in the presence of two static charges in the given
representations. When ${\cal R} = {\cal R'}=N$, the fundamental representation, 
the heavy-heavy bound state can a priori be in the adjoint or the singlet representation
(in SU(3): $ 3\otimes \bar 3 = 8 \oplus 1$). Now, it is believed that only bound states 
in the singlet representation have a finite free-energy in the confined phase. 
That means that if the heavy charges themselves are not in the singlet representation, 
virtual gluons will try and screen the chromoelectric field emanating
from this coloured bound state until it is a singlet again. Since the gluons are in the adjoint
representation, they can screen the configurations of the heavy-quark bound state that are
in the adjoint representation, albeit at a certain energy cost. 
On the other hand they cannot screen a single heavy 
quark, and the latter therefore has an infinite free energy.

Suppose we want to determine the static potential for sources in all possible
representations of SU($N$) (not necessarily irreducible) 
of a given ${\cal N}$-ality $k$ and up to a given size. Clearly
it is sufficient to determine the Polyakov loop correlators between the irreducible 
representations obtained in the decomposition of the direct product representation of $k$
quarks. The question then arises whether the cross-correlations (i.e. for 
${\cal R}\neq {\cal R'}$) between Polyakov loops in these irreducible representations
vanish or not. If they do, it implies that the energy eigenstates are in definite
irreducible representations.

Consider the $k=2$ case. The direct product of two fundamental representations 
decompose into a symmetric and an anti-symmetric representation: $N\otimes N=A\oplus S$.
In the most familiar case of SU(3), the 
anti-symmetric representation is nothing but the $\bar 3$ (anti-fundamental):
$3\otimes 3 = 6 +\bar 3$. So we are asking whether $\< P_{S}(0) P_{\bar A}(\vec x)\>$ 
has to vanish. We have $6\otimes 3 = 10 \oplus 8$, so that virtual gluons can
screen the adjoint piece, thus ensuring that the free energy of this system is finite.
So in general these cross-correlations do not vanish. It is easy to see (using Young
tableaux) that in SU($N$) the adjoint representation appears exactly once 
in the decomposition of $S\otimes \bar A$.
However since gluons have to screen the heavy-heavy system, the 
$\< P_{S}(0) P_{\bar A}(\vec x)\>$ are $1/N^2$ suppressed at large $N$.
Let us now see what this conclusion implies for the determination of the 
$k$-string tensions in the open and the closed string sectors.

In order to study the lightest open string, one may 
in principle choose to immerse any one static source of the relevant ${\cal N}$-ality
in the system, since
for large enough $R$ the sources are expected to be screened down to the 
representation with the smallest string tension. Once the linear behaviour of 
$V(R)$ with the latter slope sets in, the differences between the static potentials 
in irreducible representations of same ${\cal N}$-ality are expected to become
only weakly $R$-dependent (they correspond to `gluelump' masses~\cite{michael}). 
For long enough strings, the lowest excitations of any of these static `mesons'
correspond to the lowest excitations of that string, which come in gaps of order $1/R$.
In short, there is at most one stable open string for a given ${\cal N}$-ality.

It is also possible to interpret the Polyakov loop correlator 
with a transfer matrix along the direction $\vec x$. One is then measuring the 
spectrum of states of the gauge theory which carry a winding number with respect
to a cycle of the hypertorus of length $L$. 
Of course, since the Polyakov loop correlator has a unique asymptotic area law,
the coefficient in front of the area defines both the string tension in the
open as in the closed string sector. Just as in the open string case,
there cannot be more than one stable string per $\cal N$-ality
because of the screening by gluons. A simple picture~\cite{lucini} is that
virtual gluons screen the unstable string down to the stable one 
and propagate along it until they annihilate around the cycle of the torus.

For long enough torelons, the lowest closed-string excitations are again 
expected to be string-like, i.e. coming in $1/L$ gaps. 
There can be resonant states of the torelons 
(lying above the $k$-torelon threshold) whose energies grow linearly with $L$.
It is then natural to associate them with meta-stable strings.

What we inferred about the cross-correlations between different irreducible 
representations above tells us that the energy eigenstates do \emph{not} in 
general belong to irreducible representations of SU($N$), although 
the mixing between them is suppressed (at least in the $k=2$ case) by $1/N^2$. 
%
%
%%%%%%%%%%%%%%%
\section{Lattice simulations\label{sec:lattice}}
%%%%%%%%%%%%%%%
We extract string tensions in the three-dimensional SU(8) gauge theory
from the masses of `torelons`, gauge
invariant states transforming non-trivially under the
$Z(N)$ symmetry of the action; they wind around one spatial cycle of a
the hypertorus. These masses are extracted from the exponential
decay of correlation functions at `large' Euclidean time. 
To enhance the signal-to-noise ratio, we use fuzzing techniques
in the construction of our operators as described in~\cite{lucini04}.
The correlation
functions are measured on gauge configurations generated by a Monte-Carlo
program. We use the original Wilson action~\cite{Wilson:1974sk}.
The configuration is updated by sequences of `sweeps'.
One sweep consists of updating all links
by performing  either a heat-bath (HB)~\cite{kenpen} or an over-relaxation 
(OR)~\cite{adler} step on $N(N-1)/2$  of its $SU(2)$ subgroups~\cite{cm}. 
The ratio of HB:OR is 1:3, 
and we typically perform a sequence of 1 HB and 3 OR between measurements. 
We use a 2-level algorithm~\cite{2leva} as described in~\cite{2levb}. 
The latter reference also contains a detailed comparison of efficiency of the
 ordinary 1-level and 2-level algorithms. 
The number of measurements performed at fixed time-slices was 
800 at $\beta=115$, 200 at $\beta=138$ and 40 at $\beta=172.5$.
%%%%%%%%%%%%%%%%%%%%%%%%%%%%%%%%
\subsection{String corrections\la{sec:string_corr}}
%%%%%%%%%%%%%%%%%%%%%%%%%%%%%%%%
Consider the Euclidean gauge theory on a $L\times L\times T$
hypertorus, with cycles of length $L$. The gauge-invariant states with
winding number $k\neq 0$ around one spatial
cycle of the hypertorus are called torelons. In the Hamiltonian 
language they are created by spatial Polyakov-loop operators 
with ${\cal N}$-ality $k$; a description of the operators used 
can be found in appendix C.
If the dynamics of a torelon state  of length $L\sqrt{\sigma}\gg 1$
is described by an effective string action, then the expression
for its mass as a function of its length reads
\be
m(L)=\sigma L \left[1- \frac{\gamma}{\sigma L^2}+ {\cal O}\left(
\frac{1}{\sigma L^2}\right)^2\right], \la{eq:string_correction}
\ee
where $\gamma$ is a numerical coefficient of order one which
only depends on the universality class of the string~\cite{luscher}. 
Recent accurate numerical results~\cite{Luscher:2002qv,lucini}
show that the flux-tube 
in the fundamental representation belongs to the bosonic string class. 
In the case of a torelon this implies that 
\be
\gamma = \gamma_b \equiv \frac{(D-2)\pi}{6}. \la{eq:bosonic_value}
\ee

In general, if \eq\ref{eq:string_correction} holds,
the ratio of the lightest $k$-torelon mass to the $k=1$ torelon mass
is given by
%%%%%%%%%%%%%%%%%%%%%%%%%%%%%%%%%%%%%%%%%%%%%%%%%%%%%%%%%%%%%%%%%%%
\be
\frac{m_k}{m_1}(L)=\frac{\sigma_k}{\sigma_1}+
 \frac{\alpha_k}{\sigma_1L^2}+ {\cal O}\left(
\frac{1}{\sigma_1 L^2}\right)^2~,\qquad
\alpha_k=\frac{\gamma_1\sigma_k}{\sigma_1}-\gamma_k
\la{eq:ratio_formula}
\ee
The sign of $\alpha_k$ is of interest. If the $k$-string is a weakly bound
state of $k$ fundamental strings, then one would expect $\gamma_k=k\gamma_1$
and therefore $\alpha_k=-(k-\frac{\sigma_k}{\sigma_1})\gamma_1<0$.
If one the other hand the
fluctuations of the $k$ fundamental strings are `in phase', then
the number of degrees of freedom on the worldsheet of the $k$-string  
is the same as for the fundamental string, hence $\gamma_k=\gamma_1$
and $\alpha_k=(\frac{\sigma_k}{\sigma_1}-1)\gamma_1 > 0$.

At each lattice spacing, we measured the masses of the torelon states
of at least three different lengths. In practice,
we use two asymmetric lattices of the type $L_1\times L_2\times L_t$
and $L_2\times L_3\times L_t$. In this way, we obtain three different lengths of the 
torelon, and we can also check for any dependence on the transverse size
 of the lattice by
comparing the mass obtained for the torelon of length $L_2$ on the two lattices.
The $L_i$ range from 1.4fm to 3fm, if we set the scale by $\sqrt{\sigma_1}=440$MeV.
This is longer than what has been normally measured so far, and is made possible
by the use of the two-level algorithm. We use \eq\ref{eq:string_correction} to obtain
the fundamental string tension by fitting  $m(L)/L$ with a linear function in $1/L^2$. 
The intercept yields the string tension; the slope gives the L\"uscher coefficient. 
Whether the functional form~(\ref{eq:string_correction}) successfully describes
the leading deviation from constant linear mass density is  
controlled by the $\chi^2$ of the fit.

Systematic errors play an important role in comparing the numerical
data to model predictions. In an attempt to get them under control
we propose two separate ways to extract the ratios of string tensions
(we  refer to the first method as the `unconstrained' one, 
and the second as the `constrained' one). 
In practice, having learnt from the pros and contras of both data analyses, 
we present our final, `educated' analysis in section~\ref{sec:final_anal}.

\paragraph{1.}
Firstly the ratios of torelon masses $m_k(L)/m_1(L)$ are fitted according
to \eq\ref{eq:ratio_formula} with 
a linear function in $1/L^2$, and the intercept gives us the 
ratio $\frac{\sigma_k}{\sigma_1}$. In this way, we need make no assumption
about the values of the coefficients $\gamma$ corresponding to the 
different representations; in particular, the different strings could
have different coefficients $\gamma_k$. Finally, these string ratios 
are extrapolated to the continuum, $a\rightarrow 0$, in a standard way.

\paragraph{2.}
The second analysis will \emph{assume} that all $k$-strings
belong to the bosonic class. Consequently, we can extract the string
tension ratio from \eq\ref{eq:ratio_formula} using the estimate
$\alpha_k\simeq \frac{m_k\gamma_1}{m_1}-\gamma_k$
with $\gamma_k=\gamma_1=\gamma_b$ at every $L$. The estimates of the 
ratios obtained at different $L$ are then simply averaged, 
as long as they are compatible with eachother, to produce the estimates
of the string-tension ratios. If the $\chi^2$ of the average is large, 
we drop the smallest $L$ until an acceptable $\chi^2$ is reached. 
The continuum limit is then taken.

\paragraph{}
The multi-level algorithm allows us to 
apply the variational method~\cite{mart_var} on the correlation matrices 
at $t\geq2a$ of Euclidean time separation, an improvement
over the traditional  where the method is usually unstable unless
$t=0$, although the method really finds its justification 
when applied at large $t$. 

Having said that,
we note that this work constitutes the first attempt to extract the $k=4$ 
string tension from Monte-Carlo simulations, and should be regarded 
as exploratory in that sector. 
Indeed we found that the variational method~\cite{mart_var}
generally became unstable if all five operators listed in 
appendix C were fed in the generalised eigenvalue problem. 
As a consequence only three or four of the five types of measured operators 
(at the `best' level of smearing-blocking) were finally employed. 
This and the fact that we only have a 
short range in Euclidean time to identify the mass plateau, due to the 
rapid fall-off of the signal, means that the $k=4$ string tension 
has a significant systematic error attached to it. For the lower $k$ 
states, these problems are less accute and we are much more confident
about their mass estimates.

%%%%%%%%%%%%%%%%%%%%%%%%%%%%%%%%%%%%%%%%%%%%%%%%%%%%%
\subsection{Data analysis\la{sec:analysis}}
%%%%%%%%%%%%%%%%%%%%%%%%%%%%%%%%%%%%%%%%%%%%%%%%%%%%%
We give the masses of the lightest spatial torelons of each ${\cal N}$-ality
in \tab\ref{tab:m_T}; 
\tab\ref{tab:excited} gives estimates of the first-excited
torelon mass in the $k=2$ sector, that will be discussed below.
Within the range considered ($L\simeq1.9$fm, 
$  0.8L\leq L_\perp \leq 1.2 L$),
we certainly find no dependence of the $k=1$ torelon masses 
on the transverse size. There is also no statistically significant
variation of the lightest higher-$k$ torelon masses.
Transverse size corrections are expected to be suppressed by a power
of $1/L$ varying continuously with $L_\perp$, 
but greater than 3~\cite{Meyer:2005px}.

We show on \fig\ref{fig:loc_eff_mass} the local effective mass of the 
correlators in the $k=1$ and $k=2$ representations. 
We emphasize that the variational method, which yields (quasi-)orthogonal states,
automatically picks out the symmetric and anti-symmetric linear combinations 
(within very small fluctuations on the coefficients). We shall come back
to this point in the discussion below, section~\ref{sec:excited}.
%%%%%%%%%%%%%%%%%%%%%%%%%%%%%%%
\subsubsection{Setting the scale}
%%%%%%%%%%%%%%%%%%%%%%%%%%%%%%%
Although one could choose the (dimensionful) coupling to set the scale,
we prefer to use $\sqrt{\sigma_1}$ for this purpose. We extract the 
fundamental string tension in lattice units 
at each of our three lattice spacings by linearly extrapolating 
the torelon mass per unit length, $m_T/g^4L$ as a function of $1/(g^4L^2)$, 
to infinite $L$. This is illustrated by \fig\ref{fig:polyas} in the case
$\beta\equiv\frac{2N}{ag^2}=138$. 
The resulting string tensions are given in \tab\ref{tab:ratios}.
We are able to extract the coefficient of the $1/R$ string correction with 
moderate accuracy; it is also given in \tab\ref{tab:ratios}.
The coefficients we obtain are within 1.3 standard deviations of the bosonic
string value. 

Similarly, we can extract the $k=2$ string tension and its string correction
coefficient $\gamma_2$ (\fig\ref{fig:polyas}, bottom plot). It is clear however that the 
accuracy of the data does not allow us  to estimate $\gamma_2$.
%%%%%%%%%%%%%%%%%%%%%%%%%%%%%%%%%%%%%%%%%%
\subsubsection{Unconstrained extrapolations\la{sec:unconstrained}}
%%%%%%%%%%%%%%%%%%%%%%%%%%%%%%%%%%%%%%%%%%
In this analysis, for each lattice spacing 
we extrapolate the ratios of $k$-torelon masses to $L=\infty$, 
assuming $1/(g^2L)^2$ corrections. 
%%% This is illustrated on \fig\ref{fig:ratios_sig_u}. 
In most cases, we have three torelon lengths to extrapolate. For the 
intermediate length, where we have two statistically independent and compatible 
values obtained at different transverse sizes of the spatial lattice,
the average (weighted by the inverse square of the statistical error) of the two values
was taken, whilst keeping the smaller of the two errors. 
In the ratio of the $k$-torelon to the $k=1$ torelon mass obtained in the same 
simulation, we checked in several cases that the error bars obtained by assuming 
statistical independence do not differ by more than $10\%$ from the jacknife 
values of the error bars; the former are then used in the following.

We note that the results of these extrapolations done at different lattice spacings
are in fact consistent within error bars (see \tab\ref{tab:ratios});
it appears that finite lattice spacing effects are much smaller than the finite
string-length effects in our data set.
The $\chi^2$ of each of these fits are good (smaller than 1), 
except for the extrapolation of the $\sigma_2/\sigma_1$ at $\beta=172.5$, 
where $\chi^2=3.0$. Since the $L=\infty$ 
extrapolated value is entirely consistent with that obtained at the other values
of $\beta$, we attribute this to a statistical fluctuation and, perhaps, a slight
underestimation of the error bars (due to the neglect of the sort of systematic
errors mentioned at the end of section~\ref{sec:string_corr}).

Now extrapolating these string tension ratios 
to the continuum (assuming ${\cal O}(\sigma_1 a^2)$ discretisation errors), 
we obtain $\sigma_2/\sigma_1 = 1.701(77)$, $\sigma_3/\sigma_1 = 2.31(16)$
and $\sigma_4/\sigma_1 = 1.96(23)$.
% \ba
% \sigma_2/\sigma_1 &=& 1.701(77) \nn
% \sigma_3/\sigma_1 &=& 2.31(16)  \qquad{\rm preliminary!}     \la{eq:ratios} \\
% \sigma_4/\sigma_1 &=& 1.96(23).\nonumber 
% \ea
The $\chi^2$ of these fits are smaller than 1. The final error bars have blown
up due to a somewhat small level-arm in the continuum extrapolation.
%%%%%%%%%%%%%%%%%%%%%%%%%%%%%%%%%%
\subsubsection{Constrained analysis\la{sec:constrained}}
%%%%%%%%%%%%%%%%%%%%%%%%%%%%%%%%%%
In this independent analysis, we assume the validity of \eq\ref{eq:string_correction} 
with $\gamma$ given by the bosonic string value \eq\ref{eq:bosonic_value} to extract
the string tensions at finite $L$ (neglecting the ${\cal O}(1/L^4)$ terms); 
see the string tension ratios in \tab\ref{tab:ratios_c}, 
where again statistical errors have been added in quadrature. 
In most cases, these 
ratios are consistent with being independent of $L$ for $L\geq 1.4$fm.
The exceptions concern the $k=2$ string at the two smaller lattice
spacings (due to the accuracy of the data), 
where we drop the smallest $L$ in our average.
We note that, compared to the values of the unconstrained analysis
(\tab\ref{tab:ratios}), the ratios are systematically larger.
The ratios for the $k$-strings in the continuum limit now are:
$ \sigma_2/\sigma_1 = 1.776(33) $, $ \sigma_3/\sigma_1 = 2.210(50) $
and $ \sigma_4/\sigma_1 = 2.282(63)$.
%%%%  (\fig\ref{fig:extrapol}):
% \ba
% \sigma_2/\sigma_1 &=& 1.776(33) \nn
% \sigma_3/\sigma_1 &=& 2.210(50)   \qquad{\rm preliminary!}  \la{eq:ratios_c}\\
% \sigma_4/\sigma_1 &=& 2.282(63).     \nonumber
% \ea
The $\chi^2$ of the fits are again smaller than 1. 
%%%%%%%%%%%%%%%%%%%%%%%%%%%%%%%%%%
\subsubsection{Final `educated' analysis~\la{sec:final_anal}}
%%%%%%%%%%%%%%%%%%%%%%%%%%%%%%%%%%
We consider the preceding analysis to be somewhat unsatifactory, 
because it assumes a specific correction to the $k$-string energies
which we are not presently able to confirm directly (see \fig\ref{fig:polyas}), 
and yet (in the $k=2$ case) is of the same order of magnitude as the 
difference between two theoretical expectations we are to compare our data to.
Moreover we saw that the string tension ratios obtained in this way are systematically
higher than if we do not make any assumptions about the L\"uscher coefficients, 
although the trend is at the one-standard-deviation level.

The first analysis is well-principled but suffers from the succession 
of extrapolations to $L=\infty$  and $a=0$, most of which are based on 
three data points only and are therefore rather unstable.
Considering  the large-$L$ extrapolation (\fig\ref{fig:ratios_sig_f},
in particular the  $k=2$ plot), we see that while the coarsest lattice spacing data 
still shows a difference with respect to the other two data sets, the latter two 
essentially fall on a single curve. Therefore we drop the $\beta=115$ data and
combine the data at $\beta=138$ and $\beta=172.5$ to do a single extrapolation 
to $L=\infty$. The result is:
\ba
    \qquad\qquad     {\rm lattice~(final)} & {\rm adj. monop.} \quad {\rm fund. monop.} & {\rm trigonometric} \nn
\sigma_2/\sigma_1 = 1.707(28)     &1.714\qquad \qquad 2.105\qquad&\quad 1.848  \nn
\sigma_3/\sigma_1 = 2.182(55)     &2.143\qquad \qquad 2.958\qquad&\quad 2.414  \la{eq:final_ratio}\\
\sigma_4/\sigma_1 = 2.203(82)     &2.286\qquad \qquad 3.256\qquad&\quad 2.613  \nonumber.
\ea
(the $\chi^2$ are respectively $3.5/5$, $2.35/4$ and $2.6/4$).
One ought to associate a systematic error with this final result which is of the 
same order as the statistical error, since evidence for the absence of scaling violations
was given only at that level of accuracy. We also note that the slope, which corresponds
to the quantity $\alpha_k$ defined in section~\ref{sec:string_high_rep}, is clearly positive,
clearly demonstrating that the central charge of a $k$-string is not $k$ times
that of the fundamental string.

It is hoped that presenting different analysis strategies has given the reader a sense
of the challenge presented by these calculations to reduce the systematic errors on
the final string tension ratios.
Comparing our data to the theoretical predictions of various models
(\eq\ref{eq:final_ratio}) we find that our data
is consistent with the Casimir scaling predicted by the adjoint monopole model, 
and rules out the sine formula by at least 3 standard deviations at all $k$
(even if we conservatively assign to the data a systematic error equal to the statistical one).
These conclusions agree with earlier results obtained for SU(4) and SU(6)~\cite{lucini},
although the accuracy was high enough for $N=4$ to see a (non unexpected)
small deviation from Casimir scaling.
%%%%%%%%%%%%%%%%%%%%%%%%%%%%%%%%%%%%%%%%%%%%%%%%%%%%
\subsection{Excited $k=2$ strings\la{sec:excited}}
%%%%%%%%%%%%%%%%%%%%%%%%%%%%%%%%%%%%%%%%%%%%%%%%%%%%
\fig\ref{fig:loc_eff_mass} shows the local effective masses, defined as 
$m_{\rm eff}(t+\frac{a}{2}) \equiv \log\left(\frac{C(t)}{Ct+a)}\right)$,
of several of our operators; a plateau is the signature that
an energy eigenstate is saturating the correlator.
We show the local effective mass for our best $k=1$ operator. 
The latter has been determined by a variational method~\cite{mart_var}
allowing to minimise the contributions from excited states to the correlator.
Although several levels of fuzzing were included in the variational basis, 
the output wave function turned out to be dominated by a single level of fuzzing.
We note that its plateau extends out to $t\simeq\sigma^{-1/2}$,
giving as confidence in our mass extraction. 
In the $k=2$ sector, we show local effective masses corresponding to
the same level of fuzzing that was optimal for the $k=1$ sector (the inclusion of 
other fuzzing levels leads to imperceptible changes in the mass plateaux).
After the basis operators had been normalised in such a way that $\< O_i(0)O_i^*(t=0) \>=1~(i=1,2)$,
the variational procedure selected (within $O(1\%)$ error bars)
the anti-symmetric and the symmetric linear combinations of the operators $O_1\equiv\tr\{P^2\}$
and $O_2\equiv (\tr P)^2$ for respectively the lightest state and the first excited state.
Correspondingly these operators show quite convincing mass plateaux. 
By comparison, the individual operators have a less good overlap onto the 
lightest state, although the signal extends far enough in Euclidean time
to see that this overlap is not strongly suppressed: their local-effective-masses end up
being consistent with the plateau of the anti-symmetric combination. Remarkably
their whole correlators seem to agree at all $t$. 

Thus the theoretical expectation
that the energy eigenstates belong to irreducible representations of SU($N$)
up to $O(1/N^2)$ admixtures, which was motivated both in the two-state mixing model
and by more general arguments about the $N$-dependence of screening
(resp. sections~\ref{sec:string_high_rep} and~\ref{sec:open-closed}), 
is indeed well verified.

Another prediction of the two-state mixing model presented in section~\ref{sec:string_high_rep}
is that the lightest and the first-excited states should be split symmetrically 
around the threshold energy of $2m_{k=1}$ (to leading order in $1/N$).
This is tested quantitatively in \tab\ref{tab:excited}, which directly compares $2m_{k=1}$
to $\frac{1}{2}(m_{k=2}+m^*_{k=2})$. The latter two quantities are remarkably close
for all string lengths and lattice spacings, and in many cases they are 
compatible within the quite small error bars\footnote{As a technical aside, 
we note  that it is essential here to use the correlations between the
local effective mass of the lightest and the first-excited $k=2$ states, 
as they seem to be strongly anti-correlated.}.

As we discuss next, the numerical evidence obtained so far 
favours a binding energy of $k$-strings of order $1/N$.
The three predictions that follow straightforwardly from the two-state mixing model
presented in section~\ref{sec:string_high_rep} have thus been verified quantitatively.
%%%%%%%%%%%%%%%%%%%%%%%%%%%%%%%%%%%%%%%%%%%%%%%%
\subsection{$N$-dependence of the binding energy of $k$-strings}\label{sec:discuss}
%%%%%%%%%%%%%%%%%%%%%%%%%%%%%%%%%%%%%%%%%%%%%%%%

On \fig\ref{fig:all_n} (top)
we show the relative binding energy of $k$-strings per unit length,
$k\sigma_1-\sigma_k$, in units of $\sigma_1$ and rescaled by a factor $N$. 
We do so by compiling our SU(8) lattice 
data with the SU(4) and SU(6) data from~\cite{lucini}.
The predictions of Casimir scaling and of the Sine formula are also plotted.
The figure  certainly suggests
that the $k=2$ binding energy scales as $1/N$, 
with a coefficient of order one. By contrast, to account for the measured $N=8,~k=2$ 
binding energy in a $1/N^2$ expansion, the first coefficient would have to be about 20.
We further note that the numerical agreement between the Casimir scaling prediction
and the lattice data is quite remarkable. If anything, it lies
somewhat above the lattice data, indicating that the $k$-strings
are slightly less tightly bound that the Casimir formula suggests.

On the bottom plot, we show the $k=N/2$ string tension, rescaled by a factor $2/N$,
as a function of $1/N$. The data is plausibly heading towards a finite value at 
$N=\infty$. Here too, Casimir scaling offers a good description of the data.
Note that it predicts that the binding energy of the $k=N/2$ string is half
of the energy of $k$ non-interacting fundamental strings.
The case $k=N/2$ is special in that the relevant operators 
(listed in appendix C) and their complex conjugate can mix through the 
appearance of the baryonic vertex on the string. Pictorially it swaps the 
oriented string from on orientation to the other. Naturally the eigenstates of 
the Hamiltonian are also eigenstates of the charge conjugation operator, i.e.
the real and imaginary parts of the operators, which are respectively $C=+$ and $C=-$.
However the existence of a non-vanishing transition probability between the
strings of definite orientations means that there is a splitting between the 
$C=+$ and the $C=-$ states of ${\cal N}$-ality $N/2$ (for $k<N/2$, the 
center symmetry forces the degeneracy of these two sets of states). In a two-state
Hamiltonian formalism, the Casimir formula thus suggests that the 
Hamiltonian matrix element (per unit length of the string) 
associated with the baryon-vertex is $\frac{N}{2}\sigma$.

It should be noted that the cost of computing the binding energy for a given $k$
naively increases as $N^5$ ($N^3$ for the cost of multiplying SU($N$) together in the 
Monte-Carlo simulation and a $\propto N^2$ increase of the statistics to compensate for 
the $1/N$ size of the binding energy). And this does not even
take into account the condition
$\sigma_1 L^2\gg N$ formulated in section~\ref{sec:string_high_rep}. 
Therefore it could be useful to also 
compute the string tension ratios for SU(5) and SU(7) before moving to even larger groups.

%% file: appendix_C.tex
We give the explicit form of our Polyakov loops with ${\cal N}$-ality $k$.
Let $P(x,y,t)\equiv\prod_{n=1}^{L/a} \tilde U_x(x+na,y,t)$, where $\tilde U_x({\bf x})$
stands  either for the original link variable $U_x({\bf x})$ or a fuzzy~\cite{lucini04} 
version of it with the same gauge transformation properties. Then our operators 
are~\cite{lucini} 
\be
{\cal O}^{(k)}(t) = \frac{a}{L_y} \sum_m O^{(k)}(x,y+ma,t)  \la{eq:ops}
\ee
where the operator $O^{(k)}$ is one of
\ba
k=2: && \frac{1}{N}\tr\{P^2\}\qquad \frac{1}{N^2}(\tr\{P\})^2 \\
k=3: && \frac{1}{N}\tr\{P^3\}\qquad \frac{1}{N^3}(\tr\{P\})^3\qquad 
        \frac{1}{N^2}\tr\{P^2\}\tr\{P\} \\
k=4: && \frac{1}{N}\tr\{P^4\}\qquad \frac{1}{N^4}(\tr\{P\})^4\qquad 
        \frac{1}{N^2}(\tr\{P^2\})^2 \\
     && \frac{1}{N^2}\tr\{P^3\}\tr\{P\}\qquad \frac{1}{N^3}\tr\{P^2\}(\tr\{P\})^2, \nonumber
\ea
where the argument of $P$ is the same as that of $O^{(k)}$ in \eq\ref{eq:ops}.
The correlation functions we measure are $\< {\cal O}^{(k)}(0)({\cal O}^{(k)}(t))^*   \> $.

%% file: tab.tex
%%%%%%%%%%%%%%%%%%%%%%%%%%%%%%%%%%%%%%%%%%%%%%%%%%%%%%%%%%%%%%%%%%%%%%%%%%%%%%%%%
\clearpage
\begin{table}
\begin{center}
\begin{tabular}{|c|c|c|c|c|}
\hline %%%%%%%%%%%%%%%%%%%%%%%%%%%%%%%%%%%%%%%
I:$\beta=115.0$ & $k=1$ & $k=2$ & $k=3$ & $k=4$ \\
\small{$V=24\times28\times24$} &  &  & &  \\
\hline %%%%%%%%%%%%%%%%%%%%%%%%%%%%%%%%%%%%%%%
$L=24$ & 1.536(19)  & 2.67(13) & / &  /   \\ 
%% $L=28$ & 1.763(32)  & 3.07(46)  & /   &   /   \\
\hline %%%%%%%%%%%%%%%%%%%%%%%%%%%%%%%%%%%%%%%
\end{tabular}\\
\vspace{0.5cm}
\begin{tabular}{|c|c|c|c|c|}
\hline %%%%%%%%%%%%%%%%%%%%%%%%%%%%%%%%%%%%%%%
II:$\beta=115.0$ & $k=1$ & $k=2$ & $k=3$ & $k=4$ \\
\small{$V=16\times20\times24$} &  &  & &  \\
\hline %%%%%%%%%%%%%%%%%%%%%%%%%%%%%%%%%%%%%%%
$L=16$ & 1.0075(31)$^*$  & 1.729(30) & 2.216(66) & 2.32(12)  \\  
$L=20$ & 1.2787(33)$^*$  & 2.174(59)  & 2.55(23)  &  3.04(33)  \\
\hline %%%%%%%%%%%%%%%%%%%%%%%%%%%%%%%%%%%%%%%
\end{tabular}\\
\vspace{0.5cm}
\begin{tabular}{|c|c|c|c|c|}
\hline %%%%%%%%%%%%%%%%%%%%%%%%%%%%%%%%%%%%%%%
III:$\beta=115.0$ & $k=1$ & $k=2$ & $k=3$ & $k=4$ \\
\small{$V=12\times16\times24$} &  &  & &  \\
\hline %%%%%%%%%%%%%%%%%%%%%%%%%%%%%%%%%%%%%%%
$L=12$ &0.7338(38)$^*$  & 1.275(24) & 1.680(27) & 1.789(36) \\   
$L=16$ & 0.9980(45)$^*$ & 1.702(28) & 2.157(52) & 2.423(93)\\  
\hline %%%%%%%%%%%%%%%%%%%%%%%%%%%%%%%%%%%%%%%
\end{tabular}\\
\vspace{0.5cm}
\begin{tabular}{|c|c|c|c|c|}
\hline %%%%%%%%%%%%%%%%%%%%%%%%%%%%%%%%%%%%%%%
IV: $\beta=138.0$ & $k=1$ & $k=2$ & $k=3$ & $k=4$ \\
\small{$V=20\times24\times24$} &  &  & &  \\
\hline %%%%%%%%%%%%%%%%%%%%%%%%%%%%%%%%%%%%%%%
$L=20$ & 0.808(10) & 1.436(15) & 1.826(20)  & 1.985(29) \\
$L=24$ & 0.991(13) & 1.745(20) & 2.214(42)  &  2.28(11)  \\
\hline %%%%%%%%%%%%%%%%%%%%%%%%%%%%%%%%%%%%%%%
\end{tabular}\\
\vspace{0.5cm}
\begin{tabular}{|c|c|c|c|c|}
\hline %%%%%%%%%%%%%%%%%%%%%%%%%%%%%%%%%%%%%%%
V: $\beta=138.0$ & $k=1$ & $k=2$ & $k=3$ & $k=4$ \\
\small{$V=16\times20\times24$} &  &  & &  \\
\hline %%%%%%%%%%%%%%%%%%%%%%%%%%%%%%%%%%%%%%%
$L=16$ & 0.6348(58) & 1.1580(96) &1.464(12)  & 1.540(15) \\
$L=20$ & 0.8253(75) & 1.462(15)  &1.845(19)  & 1.937(29)  \\
\hline %%%%%%%%%%%%%%%%%%%%%%%%%%%%%%%%%%%%%%%
\end{tabular}\\
\vspace{0.5cm}
\begin{tabular}{|c|c|c|c|c|}
\hline %%%%%%%%%%%%%%%%%%%%%%%%%%%%%%%%%%%%%%%
VI:$\beta=172.5$ & $k=1$ & $k=2$ & $k=3$ & $k=4$ \\
\small{$V=24\times28\times36$} &  &  & &  \\
\hline %%%%%%%%%%%%%%%%%%%%%%%%%%%%%%%%%%%%%%%
$L=24$ & 0.5869(32) & 1.0260(82)  & 1.303(18)  &  1.401(19)   \\
$L=28$ &0.6961(58)  & 1.242(12)  & 1.564(34)  &  1.576(57)   \\
\hline %%%%%%%%%%%%%%%%%%%%%%%%%%%%%%%%%%%%%%%
\end{tabular}\\
\vspace{0.5cm}
\begin{tabular}{|c|c|c|c|c|}
\hline %%%%%%%%%%%%%%%%%%%%%%%%%%%%%%%%%%%%%%%
VII:$\beta=172.5$ & $k=1$ & $k=2$ & $k=3$ & $k=4$ \\
\small{$V=20\times24\times36$} &  &  & &  \\
\hline %%%%%%%%%%%%%%%%%%%%%%%%%%%%%%%%%%%%%%%
$L=20$ &0.4883(31) & 0.8939(65)& 1.109(12)  & 1.213(18)    \\
$L=24$ &0.5878(43) & 1.042(16) & 1.308(23)  &  1.381(22)  \\
\hline %%%%%%%%%%%%%%%%%%%%%%%%%%%%%%%%%%%%%%%
\end{tabular}
\end{center}
\caption{The masses of flux-tubes of different ${\cal N}$-alities. 
Values followed by a $^*$ were extracted from simulations employing 
a one-level algorithm. In addition, 
a $\beta=138$ run on a $32\times32\times36$ lattice
was done with $am_{k=1}=1.3347(49)$ and $am_{k=2}=2.326(49)$.}
\label{tab:m_T}
\end{table}

%%%%%%%%%%%%%%%%%%%%%%%%%%%%%%%%%%%%%%%%%%%%%%%%%%%%%%%%%%%%%%%%%%%%%%%%%%%%%%

\begin{table}
\begin{center}
\begin{tabular}{|c|c|c|c|}
\hline %%%%%%%%%%%%%%%%%%%%%%%%%%%%%%%%%%%%%%%
     & $\beta=115.0$ & $\beta=138.0$ & $\beta=172.5$  \\
\hline %%%%%%%%%%%%%%%%%%%%%%%%%%%%%%%%%%%%%%%
$a\sqrt{\sigma_1} $&0.2558(6) & 0.2059(6) & 0.1582(12) \\   %0.2557(58) &  0.2067(20)
%$\sigma_1 a^2$&  0.06545(30) & 0.04238(23) & 0.02503(38) \\ 
$\gamma_1 $     &  0.624(78)  &  0.67(12)  &  0.27(19)  \\   % 0.606(77)  & 0.77(27)
\hline%%%%%%%%%%%%%%%%%%%%%%%%%%%%%%%%%%%%%%%
$\sigma_2/\sigma_1|_{L=\infty}$ & 1.690(52) & 1.709(37) & 1.695(45) \\  % 1.702(49)old
$\sigma_3/\sigma_1|_{L=\infty}$ & 2.01(12) & 2.162(72)  & 2.175(88) \\
$\sigma_4/\sigma_1|_{L=\infty}$ & 2.28(20) & 2.31(11)   & 2.08(12) \\
\hline %%%%%%%%%%%%%%%%%%%%%%%%%%%%%%%%%%%%%%%
\end{tabular}
\end{center}
\caption{Top: the fundamental string tension and the string correction coefficient (to be 
compared with the bosonic string value $\gamma_b=\pi/6\simeq 0.5236$. Bottom:
the ratios of $k$-torelon masses, extrapolated to $L=\infty$ 
assuming $1/L^2$ corrections at three different lattice spacings. 
The continuum limit $a\rightarrow0$
of these ratios are given in section~\ref{sec:constrained}.}
\label{tab:ratios}
\end{table}

%%%%%%%%%%%%%%%%%%%%%%%%%%%%%%%%%%%%%%%%%%%%%%%%%%%%%%%%%%%%%%%%%%%%%%%%%%%%%%

\begin{table} %%%%% the constrained stuff %%%%%%%
\begin{center}
\begin{tabular}{|c|c|c|c|}
\hline %%%%%%%%%%%%%%%%%%%%%%%%%%%%%%%%%%%%%%%
$\beta=115.0$    & $k=2$ & $k=3$ & $k=4$  \\
\hline %%%%%%%%%%%%%%%%%%%%%%%%%%%%%%%%%%%%%%%
$L=12 $& 1.694(33)& 2.213(36) & 2.352(49) \\
$L=16 $& 1.688(28)& 2.141(51) & 2.319(91) \\
$L=20 $& 1.686(46)& 1.98(18)  & 2.35(26)  \\
$L=24 $& 1.728(84)&  /        &    /      \\
% $L=28 $& 1.73(26) &   /        &    /      \\
\hline%%%%%%%%%%%%%%%%%%%%%%%%%%%%%%%%%%%%%%%
Mean &  1.692(28) & 2.184(36) & 2.349(49)  \\
\hline%%%%%%%%%%%%%%%%%%%%%%%%%%%%%%%%%%%%%%%
\end{tabular}
\vspace{0.5cm}\\
\begin{tabular}{|c|c|c|c|}
\hline %%%%%%%%%%%%%%%%%%%%%%%%%%%%%%%%%%%%%%%
$\beta=138.0$    & $k=2$ & $k=3$ & $k=4$  \\
\hline %%%%%%%%%%%%%%%%%%%%%%%%%%%%%%%%%%%%%%%
$L=16 $& 1.782(22) & 2.239(28) & 2.352(32) \\
$L=20 $& 1.749(24) & 2.208(31) & 2.356(42) \\
$L=24 $& 1.744(30) & 2.207(51) & 2.27(11)  \\
$L=32 $& 1.734(37) &           &           \\
\hline%%%%%%%%%%%%%%%%%%%%%%%%%%%%%%%%%%%%%%%
Mean & 1.744(24)$^*$ & 2.222(28) & 2.349(32) \\
\hline%%%%%%%%%%%%%%%%%%%%%%%%%%%%%%%%%%%%%%%
\end{tabular}
\vspace{0.5cm}\\
\begin{tabular}{|c|c|c|c|}
\hline %%%%%%%%%%%%%%%%%%%%%%%%%%%%%%%%%%%%%%%
$\beta=172.5$    & $k=2$ & $k=3$ & $k=4$  \\
\hline %%%%%%%%%%%%%%%%%%%%%%%%%%%%%%%%%%%%%%%
$L=20 $& 1.786(18) & 2.202(28) & 2.404(44) \\
$L=24 $& 1.732(17) & 2.177(33) & 2.317(35)  \\
$L=28 $& 1.763(23) & 2.195(28) & 2.230(84)  \\
\hline%%%%%%%%%%%%%%%%%%%%%%%%%%%%%%%%%%%%%%%
Mean & 1.743(17)$^*$  & 2.195(28) & 2.304(35)$^*$  \\
\hline%%%%%%%%%%%%%%%%%%%%%%%%%%%%%%%%%%%%%%%
\end{tabular}
\end{center}
\caption{The effective ratios of $k$-string tensions,
corrected for finite-length effects assuming~\eq(\ref{eq:ratio_formula})
with bosonic string coefficient; at three different lattice spacings.
The values at different $L$
are fitted by a constant to give the 'mean' value. The mean values
were obtained by averaging values at all $L$, except for those
appearing with an asterisk $^*$, where the shortest torelon was dropped.
The continuum limit $a\rightarrow0$
of these ratios are given in section~\ref{sec:constrained}.}
\label{tab:ratios_c}
\end{table}

%%%%%%%%%%%%%%%%%%%%%%%%%%%%%%%%%%%%%%%%%%%%%%%%%%%%%%%%%%%%%%%%%%%%%%%%%%%%%%
%
%\clearpage
\begin{table}
\begin{center}
\begin{tabular}{|c|c|c|c|}
\hline
           &   $m^*_{k=2}$ & $\frac{1}{2}(m_{k=2}+m^*_{k=2})$ & $2m_{k=1}$ \\
\hline
II: $L=16$ &    2.122(60) & 1.932(33) & 2.015(6)\\ % 115.0
~~~ $L=20$&     2.49(25)  & 2.33(13) & 2.557(6)\\
\hline
III: $L=12$&    1.677(37) & 1.476(23) & 1.468(4)\\ %115.0
~~~$L=16$ &     2.244(75) & 1.973(42) & 1.996(9)\\
\hline
IV: $L=20$&     1.886(24) & 1.661(17) & 1.62(2)\\ % 138.0
~~~ $L=24$&     2.245(59)& 1.995(31) & 1.981(3)\\
\hline
V: $L=16$ &     1.380(39) & 1.272(23) & 1.27(1) \\% 138.0
~~~$L=20$ &     1.72(10)  & 1.598(54) & 1.65(2) \\
\hline
VI: $L=24$ &    1.328(15) & 1.181(9) & 1.174(6)\\ % 172.5  
~~~$L=28$ &     1.672(31) & 1.457(16) & 1.39(1) \\
\hline
VII: $L=20$&    1.109(28) & 1.002(14) & 0.977(6)\\
~~~ $L=24$&     1.296(22) & 1.174(10) & 1.176(8)\\ % 172.5  
\hline
\end{tabular}
\end{center}

\caption{The first-excited $k=2$ torelon mass. The roman numbers refer to the 
different runs whose parameters are given in \tab\ref{tab:m_T}.}
\la{tab:excited}
\end{table}
%%%%%%%%%%%%%%%%%%%%%%%%%%%%%%%%%%%%%%%%%%%%%%%%%%%%%%%%%%%%%%%%%%%%%%%%%%%%%%
%
%
% \begin{table}
% \begin{center}
% \begin{tabular}{|c|c|c|c|}
% \hline
%  & $\frac{\sigma_2}{\sigma_1}$&$\frac{\sigma_3}{\sigma_1}$&
% $\frac{\sigma_4}{\sigma_1}$\\
% \hline
% Casimir scaling & 1.714 & 2.143 & 2.286 \\
% trigonometric   & 1.848 & 2.414 & 2.613 \\
% \hline
% \end{tabular}
% \end{center}
% \caption{Casimir scaling and trigonometric formula for the ratios
% of $k$-string tensions in the SU(8) gauge theory.}
% \la{tab:expectations}
% \end{table}

%% file: fig.tex
\clearpage
%%%%%%%%%%%%%%%%%%%%%%%%%%%%%%%%% FIGURE %%%%%%%%%%%%%%%%%%%%%%%%%%%%%%%%%%%
\begin{figure}[t]
\centerline{\begin{minipage}[c]{13cm}
   \psfig{file=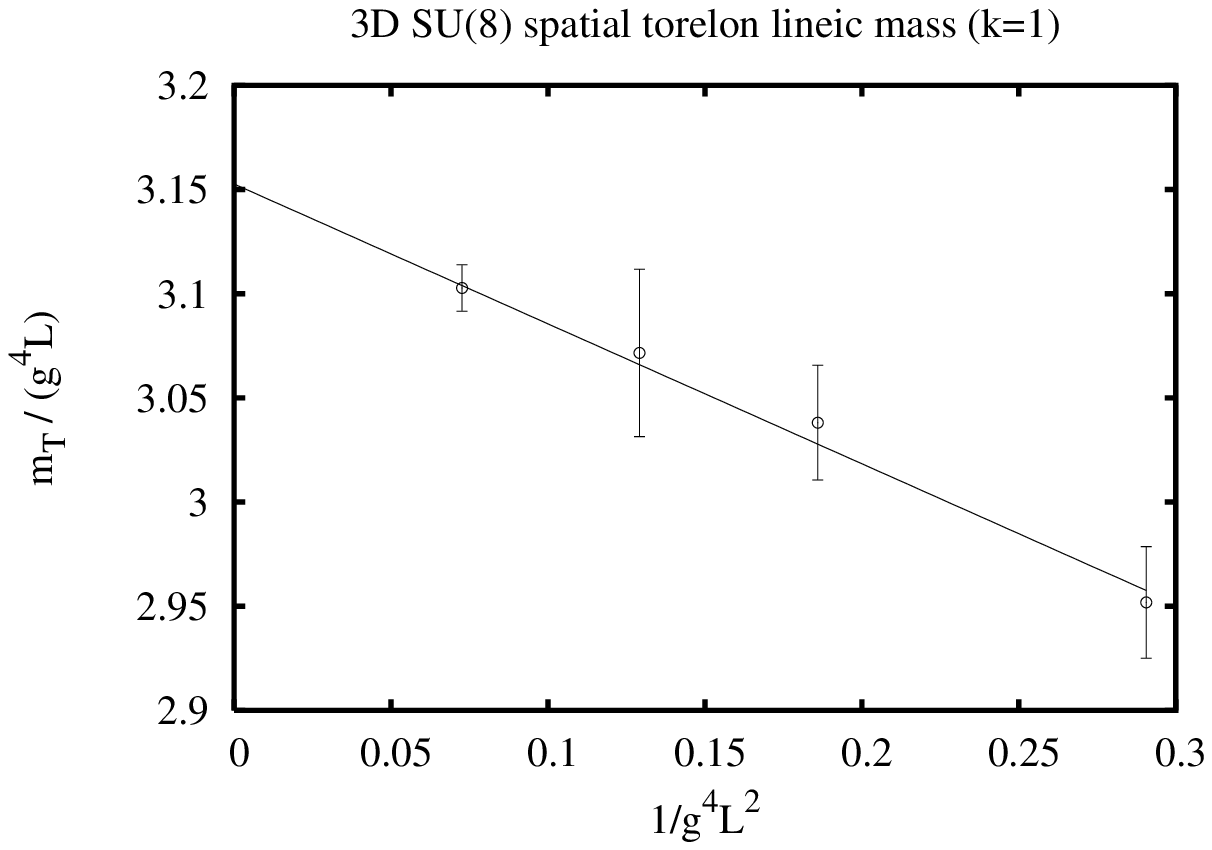,angle=0,width=13cm} 
   \psfig{file=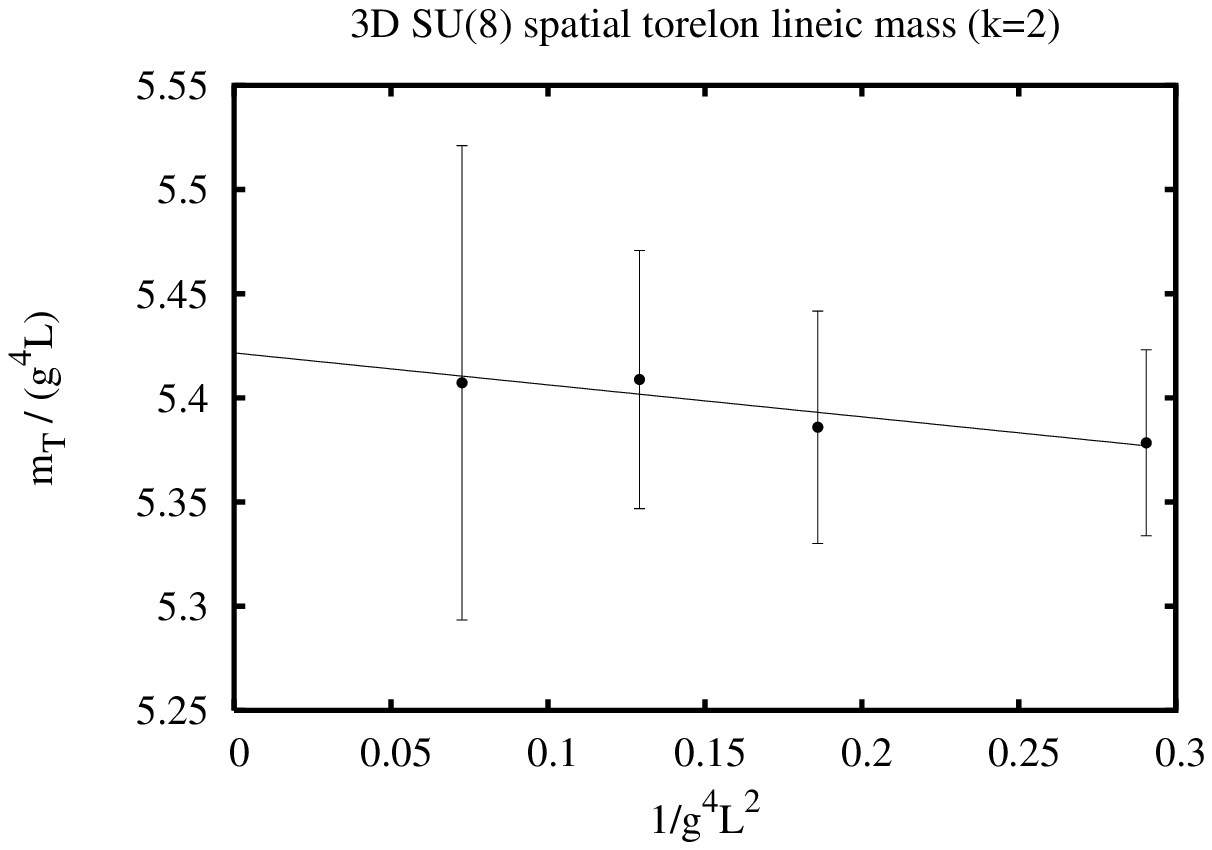,angle=0,width=13cm} 
    \end{minipage}}
\caption[a]{The lineic mass of spatial torelons in the $k=1$ and $k=2$
sectors at $\beta=138$. The intercept on the vertical axis yields the string tension.}
\la{fig:polyas}
\end{figure}
%%%%%%%%%%%%%%%%%%%%%%%%%%%%%%%%%%%%%%%%%%%%%%%%%%%%%%%%%%%%%%%%%%%%%%%%%%%%%
\clearpage
%%%%%%%%%%%%%%%%%%%%%%%%%%%%%%%%% FIGURE %%%%%%%%%%%%%%%%%%%%%%%%%%%%%%%%%%%
\vspace{-2.5cm}
\begin{figure}[t]
\centerline{\begin{minipage}[c]{10cm}
   \psfig{file=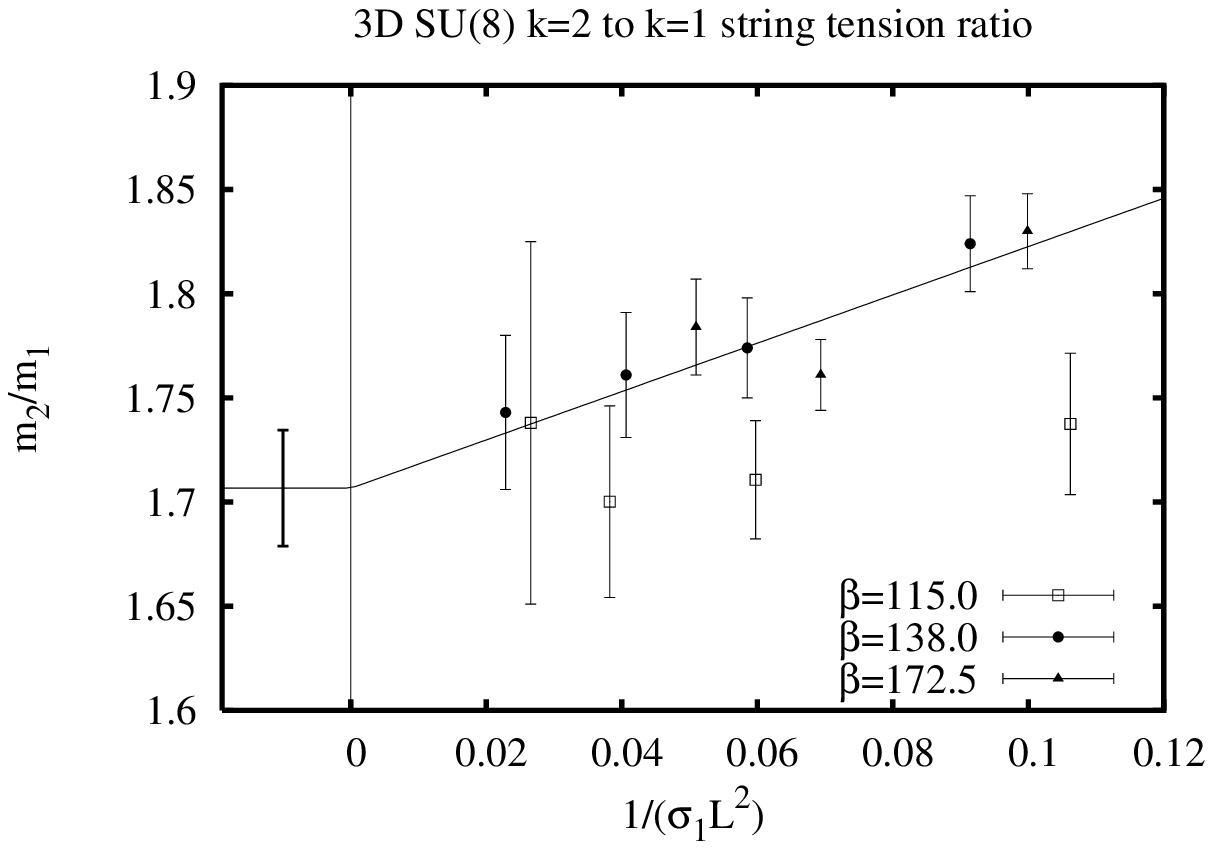,angle=0,width=10cm} \\
 \psfig{file=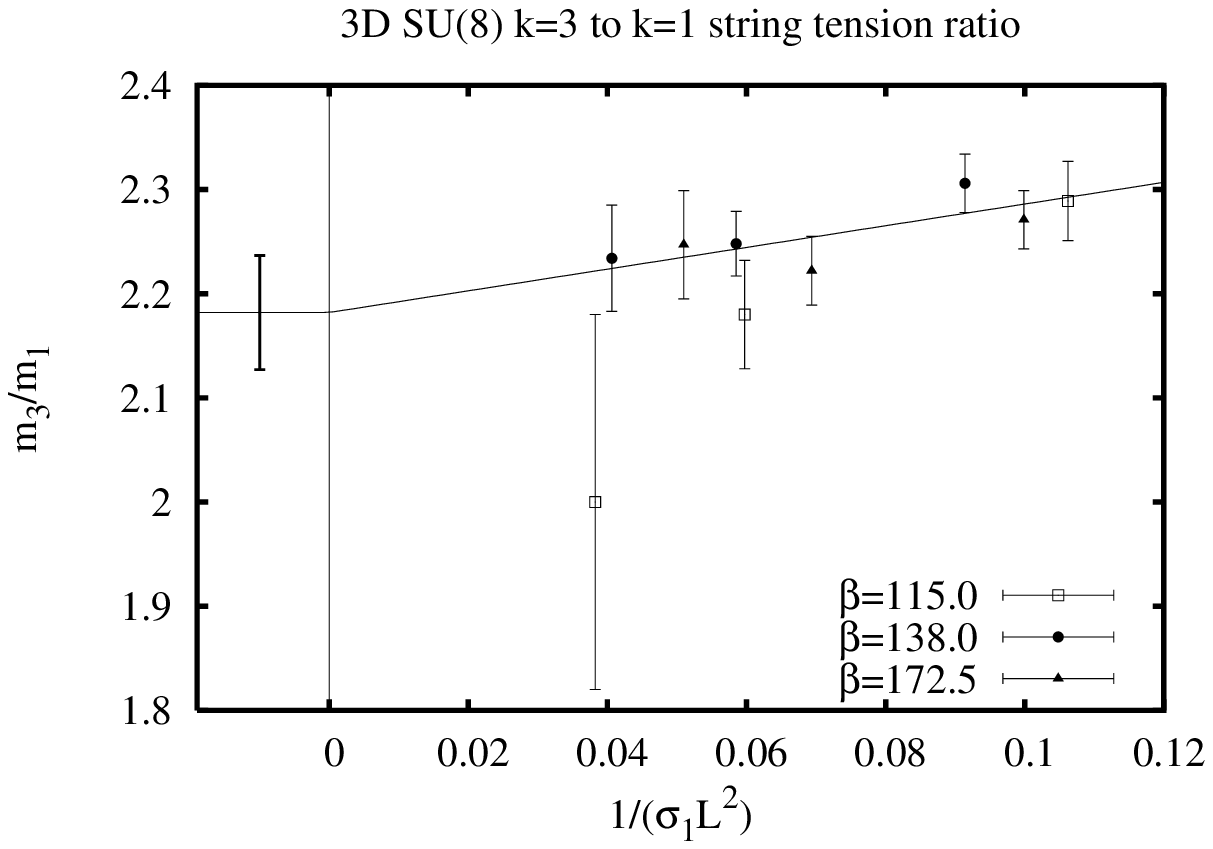,angle=0,width=10cm} \\
 \psfig{file=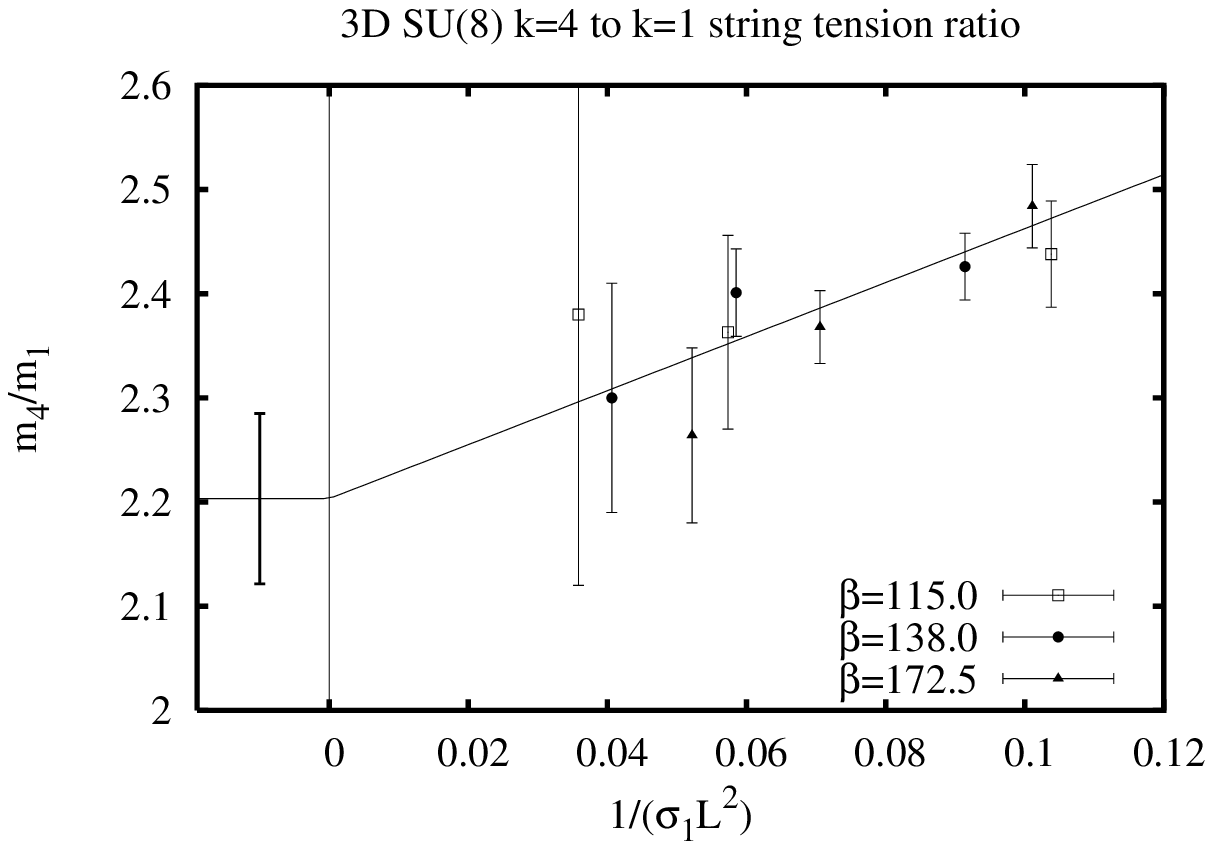,angle=0,width=10cm} 
    \end{minipage}}
\caption[a]{Final analysis: the extrapolation of the string tension ratios to infinite
string length (only the filled points are included in the extrapolation). The result
of the extrapolation is shown beyond the vertical axis.}
\la{fig:ratios_sig_f}
\end{figure}
%%%%%%%%%%%%%%%%%%%%%%%%%%%%%%%%%%%%%%%%%%%%%%%%%%%%%%%%%%%%%%%%%%%%%%%%%%%%%
\clearpage
%%%%%%%%%%%%%%%%%%%%%%%%%%%%%%%%% FIGURE %%%%%%%%%%%%%%%%%%%%%%%%%%%%%%%%%%%
\begin{figure}[t]
\centerline{\begin{minipage}[c]{12cm}
\hspace{-1.5cm}
   \psfig{file=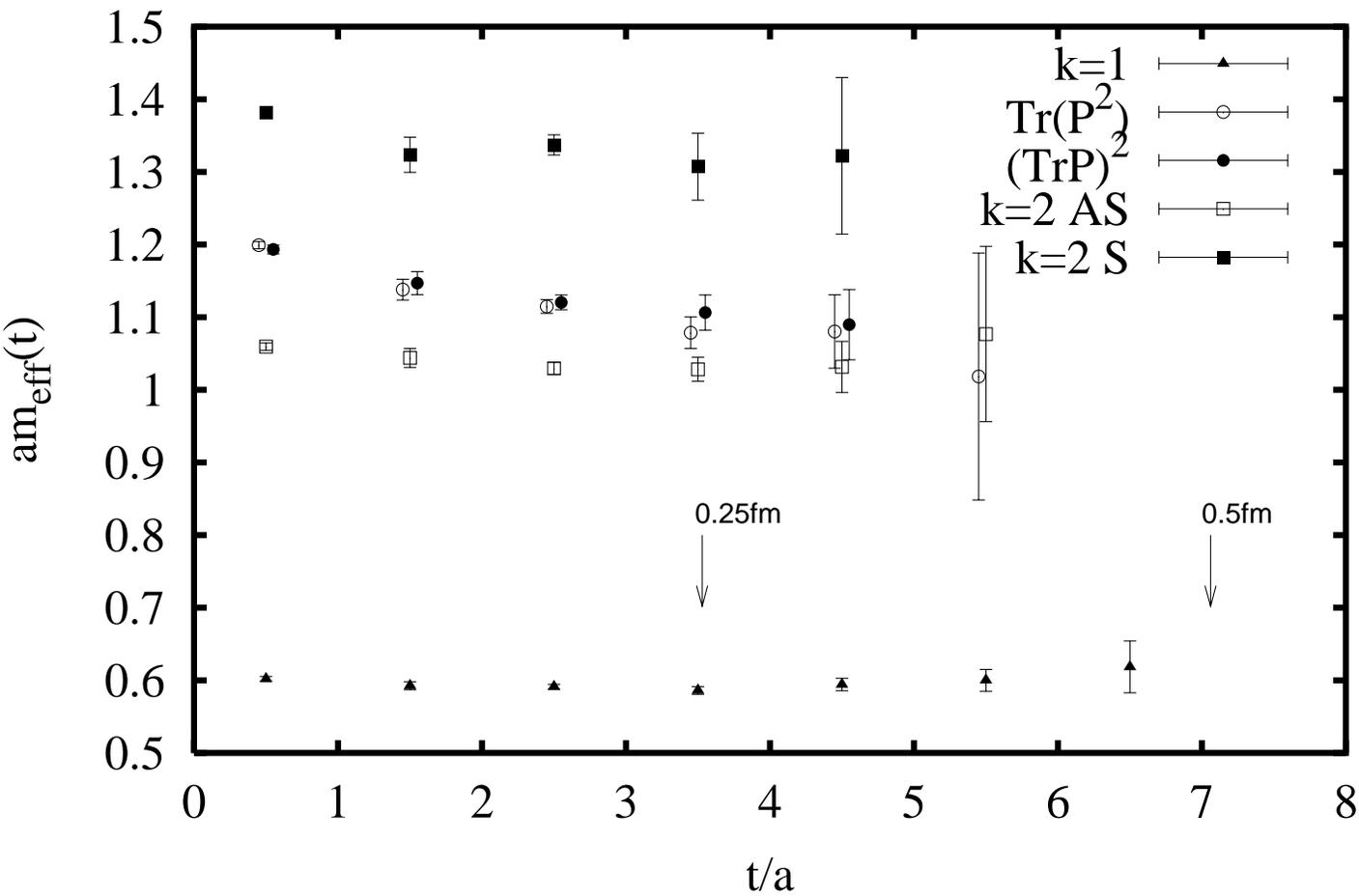,angle=90,width=14cm} 
    \end{minipage}}
\caption[a]{The  local effective mass $am_{\rm eff}(t+\frac{a}{2})
\equiv \log\left(\frac{C(t)}{Ct+a)}\right)$  for the $L=24$ k=1 and k=2 strings.
The transverse dimension is $L=28$.}
\la{fig:loc_eff_mass}
\end{figure}
%%%%%%%%%%%%%%%%%%%%%%%%%%%%%%%%% FIGURE %%%%%%%%%%%%%%%%%%%%%%%%%%%%%%%%%%%
\clearpage
\begin{figure}[t]
\centerline{\begin{minipage}[c]{12cm}
   \psfig{file=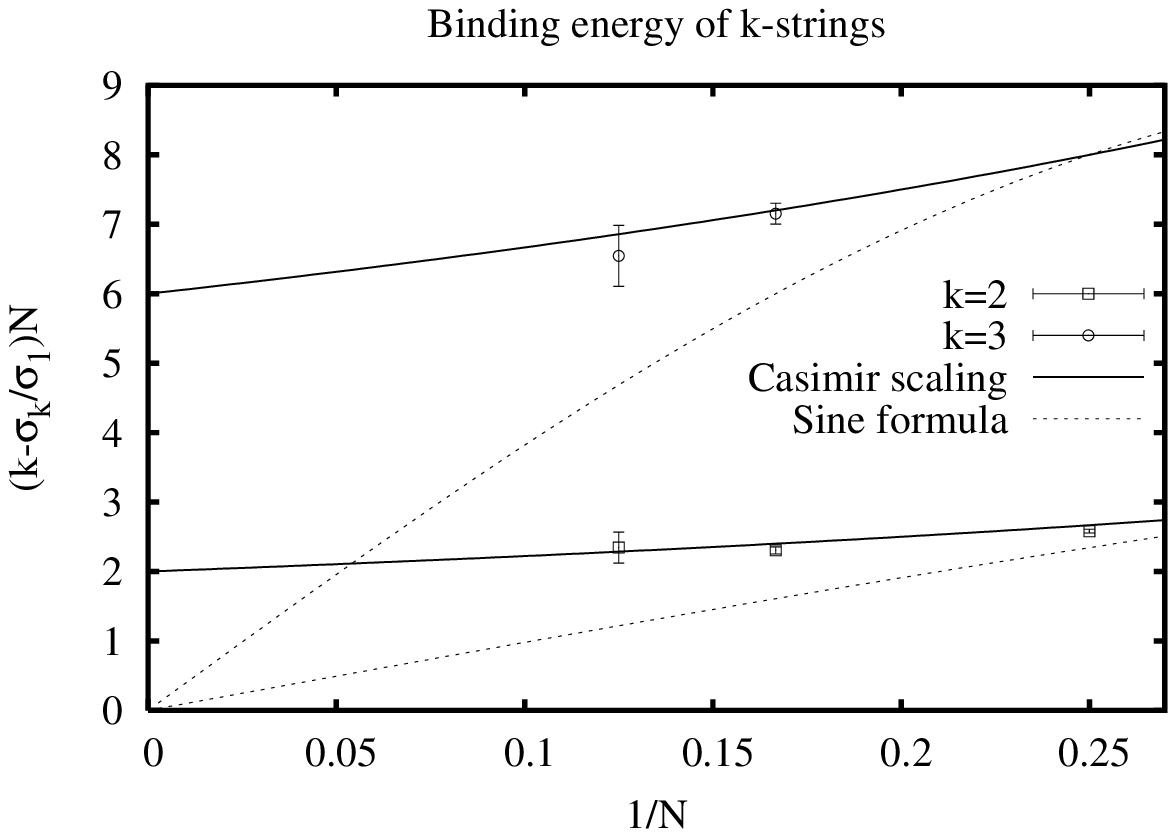,angle=0,width=12cm} 
   \psfig{file=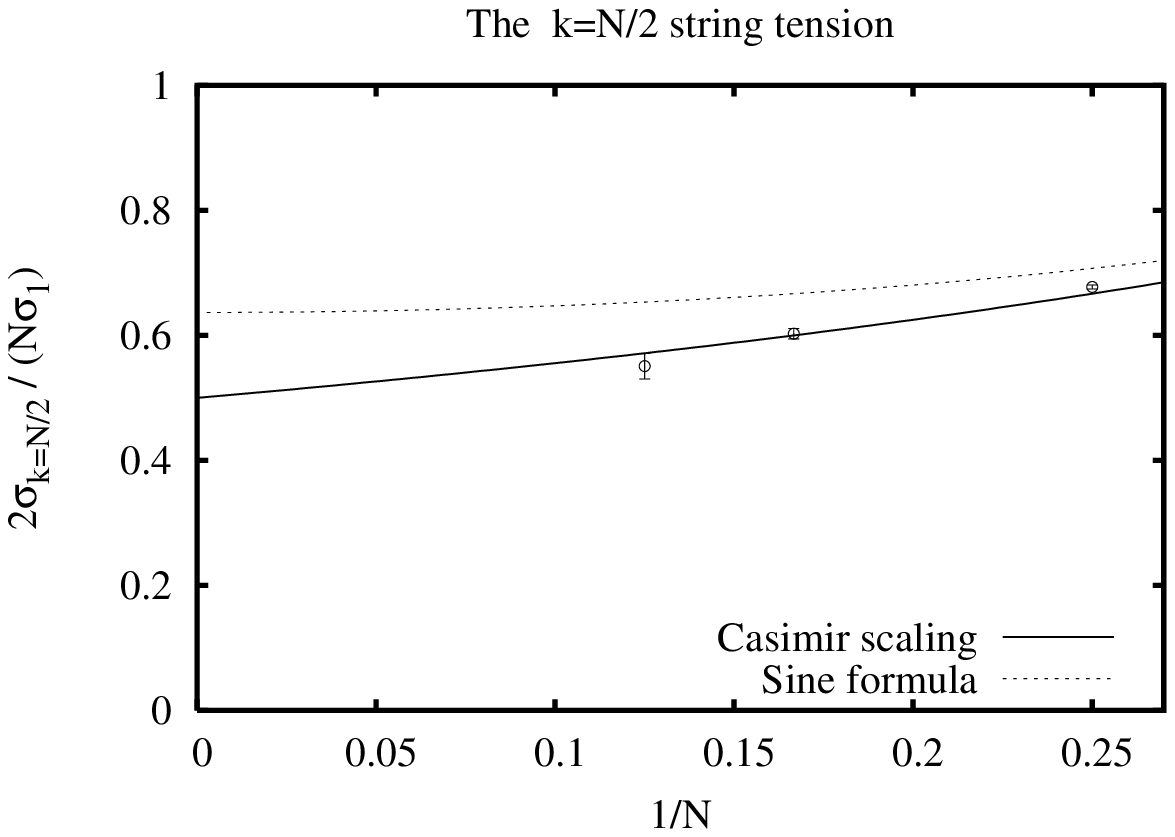,angle=0,width=12cm} 
    \end{minipage}}
\caption[a]{Top: the binding energy of $k$-strings per unit length, in units of $\sigma_1$
and rescaled by a factor $N$, as a function of $N$. Bottom: the string tension ratio
$\frac{\sigma_{k=N/2}}{\sigma_1}$, rescaled by a factor $\frac{2}{N}$.
The lattice data for SU(4) and SU(6) is taken from\cite{lucini}.}
\la{fig:all_n}
\end{figure}

%% file: papercor.bbl
\begin{thebibliography}{99}
%%%%%%%%%%%%%%%%%%%%%%
%%%%%%%%%%%%%%%%
%%%%%%%%%%%%%%%%%%%%%%%
%\cite{Luscher:1980fr}


\bibitem{tH76} G.~'t Hooft, in High Energy Physics, ed. A. Zichichi (Editrice
Compositori Bologna, 1976);
S.~Mandelstam, Phys.~Rep.~23C (1976), 245
\bibitem{th74}
 %MAGNETIC MONOPOLES IN UNIFIED GAUGE THEORIES.
 G.'t Hooft, Nucl.Phys.B79:276, 1974;  A. M. Polyakov (Landau Inst.), JETP Lett.20:194, 1974, Pisma Zh.Eksp.Teor.Fiz.20:430, 1974.

\bibitem{god}P. Goddard, J. Nuyts, D. A. Olive, Nucl. Phys.B125 (1977),
F. Englert, P. Windey, Phys. Rev.D 14(1977), 2728.
\bibitem{bais} %SPHERICALLY SYMMETRIC MONOPOLES IN NONABELIAN GAUGE THEORIES.
 F.A. Bais, J.R. Primack, Nucl.Phys.B123:253,1977; F. A. Bais, Phys.Rev.D18:1206,1978. E. J. Weinberg, Nucl. Phys. B167 (1980), 500.
\bibitem{olive80}%CHARGE QUANTIZATION IN THEORIES WITH AN ADJOINT REPRESENTATION HIGGS MECHANISM.
P. Goddard, D. I. Olive,
%ICTP/80/81-16, (Received Apr 1981). 24pp.
%Published in 
Nucl.Phys.B191, 511 (1981);
%THE MAGNETIC CHARGES OF STABLE SELFDUAL MONOPOLES.
P. Goddard, D. I. Olive,
%DAMTP-81-11, Mar 1981. 35pp.
%Published in 
Nucl.Phys.B191, 528 (1981).

%Allemaal best wel elegant met Dynkindiagrammen

 \bibitem{bais98}F. A. Bais, B. J. Schroers, Nucl. Phys. B512, 250 (1998), hep-th/9708004; Nucl.Phys.B535:197-218,1998; hep-th/9805163.
\bibitem{brandt} R.A. Brandt, F. Neri, Nucl. Phys. B 161 (1979), 253. S. Coleman, Proceedings of the 1981 Erice School, Ed. A. Zichichi,  Plenum (New York), (1982).
\bibitem{abouel}A. Abouelsaoud, Nucl.Phys.B 226 (1983), 309; P. N. Nelson, A. Manohar, Phys. Rev. Lett.50 (1983), 943; A. Balachandran et al., Phys. Rev. Lett. 50 (1983) 1553; P. N. Nelson, S.R. Coleman, Nucl.Phys.B237:1,1984; 
% NONABELIAN DUALITY IN N=4 SUPERSYMMETRIC GAUGE THEORIES.
N. Dorey, C. Fraser, T. J. Hollowood, M. A.C. Kneipp; hep-th/9512116.
\bibitem{hanany} A. Hanany, D. Tong, JHEP 0307 (2003) 037, hep-th/0306150;
% NON-ABELIAN MEISSNER EFFECT IN YANG-MILLS THEORIES AT WEAK COUPLING.%
 A. Gorsky, M. Shifman, A. Yung, Phys.Rev.D71:045010,2005, hep-th/0412082;
% NONABELIAN MONOPOLES.% R. Auzzi, S. Bolognesi, J. Evslin, K. Konoshi, A. Yung,
% Nucl. Phys.B 673,187 (2003), hep-th/0307287.




\bibitem{konishi}

 % SU(3) MONOPOLES AND THEIR FIELDS.%
 P. Irwin, Phys.Rev.D56:5200-5208,1997, hep-th/9704153;
K-M. Lee, E. J. Weinberg, P. Yi, Phys.Rev.D54:6351,1996,hep-th/9605229; 
\bibitem{humphries} J. E. Humphries, Introduction to Lie Algebras and Representation Theory, Springer, New York.
\bibitem{manuel}For a discussion of thermal screening of pointlike monopoles, see C. Manuel, Ann. Phys.263 (1998), 238.
\bibitem{linde} A.D. Linde, Phys. Lett.B 96,289 (1980).
% \bibitem{vortex}T.~G.~Kov\'acs and E.~T.~Tomboulis, Phys.~Rev.~D57 (1998) 4054;
%  T.~G.~Kov\'acs and E.~T.~Tomboulis, Phys.~Lett.~B463 (1999) 104;
% J.~M.~Cornwall, Phys.~Rev.~D57 (1998) 7589; Phys.~Rev.~D58 (1998) 1250; Phys.Rev.D65:085045,2002,  hep-th/0112230;  Phys.Rev.D71:056002,2005,  hep-ph/0412201; Phys.Rev.D70:065005,2004,  hep-th/0406084. 
%
% R.~Bertle, M.~Faber, J.~Greensite, S.~Olejnik,
%  Nucl.~Phys.~Proc.Suppl. 83 (2000) 425;
% M.~Engelhardt, K.~Langfeld, H.~Reinhardt and O.~Tennert,
% Phys.~Lett.~B431 (1998) 141.

\bibitem{polyakov} A. M. Polyakov, Nucl.Phys.B120, (1977), 429.
\bibitem{jaimungal}S. Jaimungal, G.W.  Semenoff and K.
 Zarembo, hep-th/9811238. Unpublished work by D. Diakonov and M. Chernodub (1999), private communication by D. Diakonov.
\bibitem{thooft80}G. 't Hooft, Nucl.Phys.B190, 455, 1981.

\bibitem{luscher}
M.~Luescher, K.~Symanzik and P.~Weisz,
%``Anomalies Of The Free Loop Wave Equation In The Wkb Approximation,''
Nucl.\ Phys.\ B {\bf 173} (1980) 365;  M.~Luescher,
%``Symmetry Breaking Aspects Of The Roughening Transition In Gauge Theories,''
Nucl.\ Phys.\ B {\bf 180} (1981) 317.
%%CITATION = NUPHA,B180,317;%%
%%%%%%%%%%%%%%%%%%%%%%%%%%%%%
\bibitem{bhatta}T. Bhattacharya, A. Gocksch, C.P. Korthals Altes, R. D. Pisarski, 
Nucl.Phys.B 383, (1992),497; Phys.Rev.Lett.66,998 (1991); C.P. Korthals Altes, Nucl.Phys.B420 (1994), 637.

\bibitem{giovanna01} P. Giovannangeli, C. P. Korthals Altes, Nucl. Phys.B608, 203, (2001).

\bibitem{thooftloop}  P. Giovannangeli, C. P. Korthals Altes, to appear in Nucl. Phys.B.; 
hep-ph/0412322, hep-ph/0212298. 
\bibitem{altes04}C. P. Korthals Altes, invited talk at ``Continuous Advances in QCD'', 
Minnesota, 13-16  May 2004, to appear in the  Proceedings; hep-ph/0408301. 
% \bibitem{altes03}C. P. Korthals Altes, invited talk at 
% 3rd Cracow School of Theoretical Physics: 43rd Course: Fundamental Interactions, 
% Zakopane, Poland, 30 May - 9 Jun 2003,  Acta Phys.Polon.B34:5825-5845,2003; hep-ph/0308229.
\bibitem{gliozzi}F. Gliozzi, hep-th/0504105.
\bibitem{vanbaal}% MONOPOLE CONSTITUENTS INSIDE SU(N) CALORONS.
T. C. Kraan, P. van Baal, Phys.Lett.B435:389-395,1998;  hep-th/9806034.
\bibitem{teper98}M. Teper, Phys. Rev. D59 (1999) 014512; hep-lat/9812344.
% \bibitem{Teper:1998te}
% M.~J. Teper,
% \newblock Phys. Rev. {\bf D59}, 014512 (1999), hep-lat/9804008.
\bibitem{lucini}  B.~Lucini and M.~Teper,
\newblock Phys. Rev. {\bf D64}, 105019 (2001), hep-lat/0107007.
%\cite{DelDebbio:2001sj}
\bibitem{debbio}
L.~Del Debbio, H.~Panagopoulos, P.~Rossi and E.~Vicari,
%``Spectrum of confining strings in SU(N) gauge theories,''
JHEP {\bf 0201} (2002) 009
[arXiv:hep-th/0111090].
%%CITATION = HEP-TH 0111090;%%
%%%%%%%%%%%%%%%

\bibitem{lucinidefor04}
 %MEASURING INTERFACE TENSIONS IN 4D SU(N) LATTICE GAUGE THEORIES.
 P. de Forcrand, B. Lucini, M. Vettorazzo; hep-lat/0409148

\bibitem{teper05}
% CASIMIR SCALING OF DOMAIN WALL TENSIONS IN THE DECONFINED PHA%SE OF D=3+1 SU(N) GAUGE THEORIES.
  F. Bursa, M. Teper; hep-lat/0505025.
\bibitem{lucini04}
B.~Lucini, M.~Teper and U.~Wenger, JHEP 0406:012,2004;
%``Glueballs and k-strings in SU(N) gauge theories: Calculations with improved operators,''
 hep-lat/0404008.
%%CITATION = HEP-LAT 0404008;%%
%%%%%%%%%%%%%%%%%%%%%%%%5
%%%%%%%%%%%%%%%%%
%\cite{Meyer:2002cd}
\bibitem{tepertension05}
%Properties of the deconfining phase transition in SU(N) gauge theories
%Authors: 
B. Lucini, M. Teper, U. Wenger,
%Comments: 50 pages, 14 figures
 JHEP 0502 (2005) 033; hep-lat/0502003.


\bibitem{owealtesdefor} P. de Forcrand, C.P. Korthals Altes, O. Philipsen, to appear.
\bibitem{2leva}
H.~B.~Meyer,
%``Locality and statistical error reduction on correlation functions,''
JHEP {\bf 0301} (2003) 048, hep-lat/0209145.
%%CITATION = HEP-LAT 0209145;%%

\bibitem{2levb}
H.~B.~Meyer,
%``The Yang-Mills spectrum from a 2-level algorithm,''
JHEP {\bf 0401} (2004) 030, hep-lat/0312034.
%%CITATION = HEP-LAT 0312034;%%
%%%%%%%%%%%%%%%%%%%%%%%%%%%%%
%\cite{Meyer:2003hy}
\bibitem{finivol}
 %THE SPECTRUM OF SU(N) GAUGE THEORIES IN FINITE VOLUME.
 H.~B.~Meyer, JHEP 0503:064,2005;  hep-lat/0412021.

%%%%%%%%%%%%%%%%%%%%%%%%%%%%%
%\cite{Armoni:2003ji}
\bibitem{pis} T. Applequist, R.D. Pisarski, Phys.Rev.D23,2305,(1981); P. Ginsparg, Nucl.Phys.B170,388, (1980). E.~Braaten and A.~Nieto,
%``Effective field theory approach to high temperature thermodynamics,''
Phys.\ Rev.\ D {51} (1995) 6990
;hep-ph/9501375. 
%3-D PHYSICS AND THE ELECTROWEAK PHASE TRANSITION: PERTURBATION THEORY.
 K. Farakos, K. Kajantie, K. Rummukainen, M. E. Shaposhnikov, Nucl.Phys.B425:67,  1994;  hep-ph/9404201.
%%CITATION = HEP-PH 9501375;%%

%``Free Energy of QCD at High Temperature,''
\bibitem{huang}S. Z. Huang, M. Lissia, Nucl.Phys.B438,54,1995; hep-ph/9411293.
\bibitem{kaj1997} %SU(N)JOINT HIGGS THEORY AND FINITE TEMPERATURE QCD.
 K. Kajantie, M. Laine, K. Rummukainen, M. Shaposhnikov,  Nucl.Phys.B503:357;  hep-ph/9704416.                                                                                                                               
\bibitem{yorkmikko}
 M. Laine, Y. Schroeder, hep-ph/0503061.

\bibitem{armoni}
A.~Armoni and M.~Shifman,
%``On k-string tensions and domain walls in N = 1 gluodynamics,''
Nucl.\ Phys.\ B {\bf 664} (2003) 233, hep-th/0304127;
%%CITATION = HEP-TH 0304127;%%
%%%%%%%%%%%%%%%%%%%%%%%%%%%%%%
A.~Armoni and M.~Shifman,
Nucl. Phys.B 671.67 (2003), hep-th/0307020.
\bibitem{kleb}A. Hanany, M.J. Strassler, A. Zaffaroni, Nucl.Phys B513, 87 (1998), hep-th/9707244; C.P. Herzog, I. R. Klebanov, Phys. Lett.B526, 388 (2002), hep-th/0111078.
\bibitem{kovneraltes}C. P. Korthals Altes, A. Kovner,
Phys. Rev D62, 096008, 2000; hep-ph/0004052.
\bibitem{herzog}C. P. Herzog, Phys. Rev.D 66, 065009; hep-th/0205064.
\bibitem{washington}D. J. Gross, W. Taylor, Nucl.Phys.B403:395,1993; hep-th/9303046 . 
\bibitem{bronoff} S. Bronoff, C. P. Korthals Altes,  Phys.Lett.B448:85, 1999; hep-ph/9811243.
\bibitem{kaj98}K. Kajantie, M. Laine, A. Rajantie, K. Rummukainen, M. Tsypin, JHEP 9811:011,1998; hep-lat/9811004.
\bibitem{rajantie}A. Rajantie, Nucl. Phys.B 501, 521; hep-ph/9702255.

\bibitem{rajantiealtes} C.P. Korthals altes, A. Rajantie, in preparation.

\bibitem{giovanna03}
% TWO LOOP RENORMALIZATION OF THE MAGNETIC COUPLING IN HOT QCD.
 P. Giovannangeli, Phys.Lett.B585:144, 2004; hep-ph/0312307; hep-ph/0506318.

%\bibitem{giovanna05} P. Giovannangeli, H. Meyer and C.P. Korthals
%Altes, in preparation.

\bibitem{diakonov}
%  A FORMULA FOR THE WILSON LOOP.
 D. Diakonov, V.Yu. Petrov, Phys.Lett.B224:131, 1989;
D.~Diakonov and V.~Petrov,
  %``Non-Abelian Stokes theorems in Yang-Mills and gravity theories,''
  J.\ Exp.\ Theor.\ Phys.\  {\bf 92} (2001) 905
  [arXiv:hep-th/0008035].
  %%CITATION = HEP-TH 0008035;%%
\bibitem{gubarev}
  F.~V.~Gubarev,
  %``On the non-Abelian Stokes theorem for SU(2) gauge fields,''
  Phys.\ Rev.\ D {\bf 69} (2004) 114502
  [arXiv:hep-lat/0309023].
  %%CITATION = HEP-LAT 0309023;%%
B.~Broda,
  %``Non-Abelian Stokes theorem in action,''
  arXiv:math-ph/0012035.
  %%CITATION = MATH-PH 0012035;%%
D.~Diakonov and V.~Petrov,
  %``Non-Abelian Stokes theorems in Yang-Mills and gravity theories,''
  J.\ Exp.\ Theor.\ Phys.\  {\bf 92} (2001) 905
  [arXiv:hep-th/0008035].
%%%%%%%%%%%%%%%%%%%%%%%%%%%%%%
%\cite{'tHooft:1973jz}
\bibitem{hooft}
G.~'t Hooft,
%``A Planar Diagram Theory For Strong Interactions,''
Nucl.\ Phys.\ B {\bf 72} (1974) 461.
%%CITATION = NUPHA,B72,461;%%

\bibitem{witten_baryons}
E.~Witten,
\newblock Nucl. Phys. {\bf B160}, 57 (1979).

\bibitem{Deldar:1999vi}
S.~Deldar,
\newblock Phys. Rev. {\bf D62}, 034509 (2000), hep-lat/9911008;
%\bibitem{Bali:2000un}
G.~S. Bali,
\newblock Phys. Rev. {\bf D62}, 114503 (2000), hep-lat/0006022.

%\cite{Campbell:1985kp}
\bibitem{michael}
N.~A.~Campbell, I.~H.~Jorysz and C.~Michael,
%``The Adjoint Source Potential In SU(3) Lattice Gauge Theory,''
Phys.\ Lett.\ B {\bf 167} (1986) 91.
%%CITATION = PHLTA,B167,91;%%

\bibitem{Close:2002zu}
F.~E.~Close and N.~A.~Tornqvist,
J.\ Phys.\ G {\bf 28} (2002) R249
[arXiv:hep-ph/0204205].

%\cite{Das:1984nb}
\bibitem{Das:1984nb}
S.~R.~Das,
%``Some Aspects Of Large N Theories,''
Rev.\ Mod.\ Phys.\  {\bf 59} (1987) 235.
%%CITATION = RMPHA,59,235;%%

\bibitem{Luscher:2002qv}
M.~Luescher and P.~Weisz,
\newblock JHEP {\bf 07}, 049 (2002), hep-lat/0207003.

%\cite{Meyer:2005px}
\bibitem{Meyer:2005px}
H.~B.~Meyer,
%``Vortices on the worldsheet of the QCD string,''
arXiv:hep-th/0506034.
%%CITATION = HEP-TH 0506034;%%
%%%%%%%%%%%%%%%%%%%%%%%%%%%%%
\bibitem{Wilson:1974sk}
K.~G.~Wilson,
%``Confinement Of Quarks,''
Phys.\ Rev.\ D {\bf 10} (1974) 2445.
%%CITATION = PHRVA,D10,2445;%%
%%%%%%%%%%%%%%%%%%%%%%%%%%%%%
%\cite{Albanese:ds}
%\bibitem{smear}
%M.~Albanese {\it et al.}  [APE Collaboration],
%%``Glueball Masses And String Tension In Lattice QCD,''
%Phys.\ Lett.\ B {\bf 192} (1987) 163.
%%%CITATION = PHLTA,B192,163;%%
%%%%%%%%%%%%%%%%%%%%%%%%
%\cite{Teper:wt}
%\bibitem{block}
%M.~Teper,
%%``An Improved Method For Lattice Glueball Calculations,''
%Phys.\ Lett.\ B {\bf 183} (1987) 345.
%%%CITATION = PHLTA,B183,345;%%
%%%%%%%%%%%
\bibitem{mart_var}
M.~Luscher and U.~Wolff,
\newblock Nucl. Phys. {\bf B339}, 222 (1990).
%%%%%%%%%%%
\bibitem{cm} N.~Cabibbo, E.~Marinari, Phys. Lett. B119(1982) 387
%%%%%%%%%%%
\bibitem{kenpen}K.~Fabricius, O.~Haan, Phys. Lett B143 (1984) 459;\\
 A.D.~Kennedy, B.J.~Pendleton, Phys.~Lett., 156B (1985) 393
%%%%%%%%%%
\bibitem{adler}S.L.~Adler, Phys. Rev. D 23 (1981) 2901
%%%%%%%%%%%%%
\end{thebibliography}
